\documentclass[12pt,preprint]{aastex}

\newcommand\kms{km s$^{-1}$}
\newcommand\msun{\ifmmode{M_{\odot}}\else $M_{\odot}$\fi}
\newcommand\rsun{\ifmmode{R_{\odot}}\else $R_{\odot}$\fi}

\begin{document}

\title{Elemental Abundance Ratios in Stars of the Outer Galactic 
Disk.\ II.\ Field Red Giants\footnote{This
paper makes use of obsevations obtained at the National Optical Astronomy
Observatory, which is operated by AURA, Inc., under contract from
the National Science Foundation. We also employ data products
from the Two Micron All Sky Survey, which is a joint project
of the University of Massachusetts and the Infrared Processing
and Analysis Center/California Institute of Technology, funded by the
National Aeronautics and Space Administration and the National
Science Foundation}}

\author{Bruce W.\ Carney}
\affil{Department of Physics \& Astronomy, University of North
Carolina, Chapel Hill, NC 27599-3255; email: bruce@physics.unc.edu}

\author{David Yong}
\affil{Department of Physics \& Astronomy, University of North
Carolina, Chapel Hill, NC 27599-3255; email: yong@physics.unc.edu}

\author{Maria Lu\'{i}sa Teixera de Almeida}
\affil{Department of Physics \& Astronomy, University of North
Carolina, Chapel Hill, NC 27599-3255; email: luisa@oal.ul.pt}

\author{Patrick Seitzer}
\affil{Department of Astronomy, 818~Dennison Building, 
University of Michigan,
Ann Arbor, MI 48109; email: pseitzer@umich.edu}

\begin{abstract}
We summarize a selection process to identify red giants in
the direction of the southern warp of the Galactic disk, employing
$VI_{C}$ photometry and multi-object spectroscopy. We also present
results from follow-up high-resolution, high-S/N echelle spectroscopy
of three field red giants,
finding [Fe/H] values of about $-0.5$.
The field stars, with Galactocentric distances 
estimated at 10 to 15 kpc, support the
conclusion of Yong, Carney, \& de~Almeida (2005) that
the Galactic metallicity gradient disappears beyond $R_{\rm GC}$ values
of 10 to 12~kpc for the older stars and clusters of the outer disk.
We summarize the detailed abundance patterns for 15 other 
elements for these stars and
compare them to recently-obtained results for old open cluster
red giants in the outer disk. 
The field and cluster stars at such large distances show
very similar abundance patterns, and, in particular, all
show enhancements of the ``$\alpha$" elements O, Mg, Si, Ca, and Ti
and the $r$-process element Eu. These results suggest that
Type~II supernovae have been significant contributors to
star formation in the outer disk 
relative to Type~Ia supernovae within the past few Gyrs.
We also compare our results 
with those available for much younger objects.
The limited
results for the H~II regions and B~stars in the outer disk also suggest
that the radial metallicity gradient in the outer
disk is shallow or absent.
The much more extensive
results for Cepheids confirm these trends, and
that the change in slope of the metallicity gradient
may occur at a larger Galactocentric distance than
for the older stars and clusters. However, the younger stars also show
rising $\alpha$ element enhancements with increasing $R_{\rm GC}$,
at least beyond 12~kpc.
These trends are 
consistent with the idea of a progressive growth in the
size of the Galactic disk with time, and episodic
enrichment by Type~II supernovae as part of the disk's growth.

\end{abstract}


\section{INTRODUCTION}

In this paper we continue our discussion of metallicities, [Fe/H],
and elemental abundance ratios, [X/Fe], for stars in the outer
Galactic disk. In Paper~I (Yong, Carney, \& de~Almeida 2005), we described a
set of radial velocity measures and abundance determinations of four old
open clusters whose Galactocentric distances, $R_{GC}$, lie between 12
and 23 kpc. Three stars in the local old open cluster M67 were observed
and analyzed in the same fashion to provide checks on both the
velocities and abundances.

Here we investigate a slightly different sample, three field red
giants. Selecting distant field giants in the sea of foreground
field stars, and in the presence of large and variable
interstellar extinction, presents a challenge, but
we have exploited a few key tricks. First, we believe that the
distant outer disk is detectable, based on our earlier work in
two key directions toward the southern hemisphere Galactic
warp (Carney \& Seitzer 1993). Second, modulo typical disk
population line-of-sight velocity dispersions ($\approx 20$ \kms), disk stars
obey a fairly clear rotation curve. Depending on the
Galactic longitude, radial velocities of disk stars
closer to and farther from the Sun will yield rather different
radial velocities. We exploit this here by first obtaining
extensive $VI_{C}$ photometry in the general direction of one
of the fields studied by Carney \& Seitzer. From the color-magnitude
diagram, we identified likely red giants. We then obtained high-resolution,
low signal-to-noise (S/N) spectra of over two hundred of these stars,
and identified those whose radial velocities are most consistent
with large Galactocentric distance. Three of these stars provide the
program sample for this paper.

\section{OBSERVATIONS}

The disk of the Milky Way is warped, as seen using
H~I gas (Henderson et al.\ 1982;
Burton \& Te~Lintel Hekkert 1986; Diplas \& Savage 1991),
CO emission from IRAS point sources (Wouterloot et al.\ 1990),
and the dust layer, as seen by IRAS (Sodroski et al.\ 1987)
and DIRBE (Freudenreich et al.\ 1994). Unlike many other
disk galaxies, the Milky Way's warp appears to contain
stars, including OB stars (Miyamoto et al.\ 1988; Orsatti 1992;
Reed 1996), dusty-shell stars selected from the IRAS point source
catalog (Djorgovski \& Sosin 1989), and K giants (L\'{o}pez-Corredoira
et al.\ 2002). 

Detecting stars at large distances along lines of sight
confined to the plane is nearly hopeless (except at infrared
wavelength) due to interstellar extinction. Even methods
that rely on gravity-sensitive intermediate-band photometric
features, like the $DDO$-51 filter, are vulnerable to
changes in reddening. But the warp
offers us a possible means to peek above or below the
Galactic murk and discern the most distant stars.
Carney \& Seitzer (1993) detected excess numbers of
stars along lines of sight to the southern Galactic warp
compared to lines of sight at the same Galactic longitudes
but opposite latitudes.

If we assume pure circular rotation for distant disk
stars, the radial velocity of a star at Galactic
longitude ($\ell$) and latitude ($b$) to be
\begin{equation}
\label{eq:vrad}
V_{LSR} = (\frac{\Theta}{R_{\rm GC}} - \frac{\Theta_{0}}{R_{0}}) R_{0} 
\sin~\ell \cos~b,
\end{equation}
where $R_{0}$ is the distance of the Sun from the Galactic center, $\Theta$
is the linear rotation speed of the Galactic disk at r=$R_{GC}$, and
$\Theta_{0}$ is the linear rotation speed at $R_{0}$. We adopt
$R_{0}$ = 8 kpc, $\Theta_{0}$ = 220 \kms, and that the rotation curve
beyond the solar circle is flat, 
so $\Theta$ is also 220 \kms. Table~\ref{tab:warp}
summarizes predicted purely circular, flat rotation curve values for
$V_{LSR}$ as a function of $R_{GC}$, in the line of sight for one
of the fields observed by Carney \& Seitzer (1993: their field ``Warp 1a"), 
centered at
$\ell$ = 245.75 and $b$ = $-4.1$.
Of course, we do not measure $V_{\rm LSR}$, but the heliocentric
radial velocity, $V_{\rm rad}$.
The conversion of 
$V_{\rm rad}$, to $V_{\rm LSR}$ was determined using 
Equation~\ref{eq:vlsr}, where we assumed a Solar
peculiar velocity, $V_{\rm p}$, of 20 \kms\ directed
toward $\ell_{\rm p}$ = 57 and $b_{\rm p} = 22$. 
\begin{equation}
\label{eq:vlsr}
V_{\rm LSR} = V_{\rm rad} + V_{\rm p} \times
\frac{(\cos~b_{\rm p}~\cos(\ell - \ell_{\rm p}) \cos~b + \sin~b_{\rm p} \sin~b)}{\cos~b}
\end{equation}

\subsection{Photometry}

We observed Warp Field 1a with the University of Michigan's
Curtis Schmidt telescope
at Cerro Tololo Inter-American Observatory using a Thompson 
$1024 \times 1024$ Thompson CCD
with an image scale of 1.835 arcsec per pixel 
and a field of view of 31\arcmin\ $\times$
31 \arcmin. 
The CCD was a thick frontside-illuminated device
and coated to provide $U$ and $B$ sensitivity. We obtained
images in $B$, $V$, and $I$, with exposure times ranging
from 30 to 300 seconds. After using DAOPHOT to measure
aperture magnitudes, we employed mean extinction coefficients
and a set of 16 photometric standards from Graham (1982)
to complete the transformation of the data to the standard
Cousins photometric system. The scatter in the transformations
was about 0.025 mag per star, which is reasonably good considering
the pixel scale of the data.

\begin{figure}
\plotone{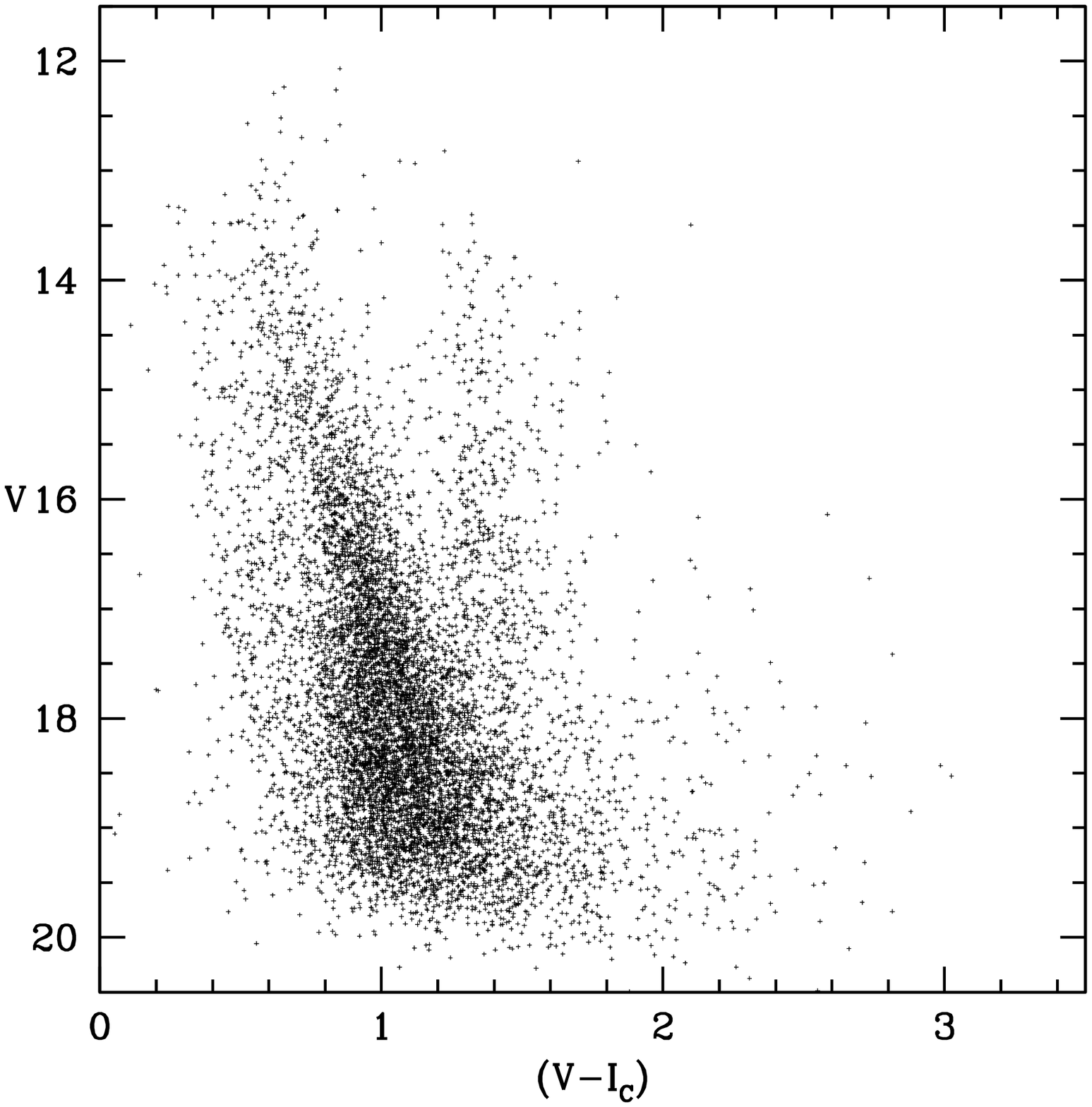}
\caption{The color-magnitude diagram in the direction of Warp Field 1a
(following Carney \& Seitzer (1993), covering a field of view of
roughly 31 arcminutes. \label{fig:cmd}}
\end{figure}

Table~\ref{tab:warpphot} summarizes our results for 8472 stars, and the
resulting color-magnitude diagram is shown in Figure~\ref{fig:cmd}.
It reveals a relatively
typical Galactic disk population of field stars, and we assume
that the stars with 12.5 $\leq\ V \leq\ 17$ and $1.2 \leq\ V-I_{C} \leq\ 1.8$
are probably red giants lying over a wide (and indeterminate) range
of distances. Our challenge was to select a few such stars for
follow-up spectroscopic studies to determine [Fe/H] and chemical
abundance patterns, and we decided to employ Equation~\ref{eq:vrad} to do so.
Therefore, we needed to obtain radial velocities for a large number
of those candidate red giants.

\subsection{Spectroscopic Observations: Multi-object Spectroscopy}

From the color-magnitude and color-diagrams constructed using
our photometry, we selected stars for radial velocity measurements.
Only stars redder than $V-I_{C}$ = +1.0 were selected.
The 24-fiber multi-object Argus echelle spectrograph was used at the
prime focus of the Cerro Tololo Inter-American Observatory 4.0-m Blanco
Telescope in March of 1994. Single-order spectra centered on the
Mg~$b$ triplet ($\approx$\ 5200 \AA) were obtained for 209 of the
stars that appear to be field red giants. We also obtained
spectra for a number of well-observed red giants in the globular
cluster $\omega$~Cen as well as bright radial velocity standards,
HD~31871, HD~120223, and HD~176047.
Radial velocities were determined using cross correlation techniques,
and the typical velocity precision for our program stars is 
2.5 \kms.
Table~\ref{tab:argusvels} summarizes our results, along with
$VI_{C}$ photometry from Table~\ref{tab:warpphot}, and
$JHK$ photometry 
from 2MASS\footnote{The
Two Micron All-Sky Survey is a joint project of the
University of Massachusetts and the Infrared Processing and
Analysis Center/California Institute of Technology, funded by
the National Aeronautics and Space Administration and the
National Science Foundation.}. 
The positions have been taken from 2MASS.

\subsection{Selection of Targets}

We selected three candidate red giants for further study. 
The red giant candidates in Figure~\ref{fig:cmd} are faint,
and we were restricted to the brighter stars. We therefore
employed the radial velocities from Table~\ref{tab:argusvels}
in comparison with the Galactic rotation model predictions
from Table~\ref{tab:warp}. All three stars have large
radial velocities and relatively bright $V$ magnitudes.
Figure~\ref{fig:chart} provides finding charts
for the three stars.

\subsection{Spectroscopic Observations: Echelle}

The three targets were observed with
the echelle spectrograph and the 4-meter
telescope at
the Cerro Tololo Inter-American Observatory (CTIO) during
January 1998 and 1999. We employed the
long red camera and the 31.6 lines~mm$^{-1}$
echelle grating. A GG495 filter blocked
second-order blue light, while the G181
cross disperser 
(316 lines~mm$^{-1}$) led to a
wavelength coverage of
5200-7940 \AA. The
slit was opened to 150 microns, providing a width of
1.0\arcsec\ on the sky, yielding a spectral
resolving power of 28,000 and a dispersion of 0.07 \AA\ per pixel
at 5800~\AA, and providing two pixels per resolution element.
Our goal was to derive accurate metallicities, [Fe/H],
as well as detailed element-to-iron abundance ratios
to explore the chemical evolution of outer disk stars,
following the same goals as presented in our previous
paper (Yong et al.\ 2005).

The observing routine included 20 quartz lamp exposures
to provide data for flat-fielding, and 15 zero-second
exposures (to provide ``bias" frames). Th-Ar hollow
cathode lamp spectra were taken before and after each
stellar exposure. For radial velocity standards, we
used the sky, and also HD~80170, which is a K giant,
and similar
in temperature and gravity to our
program stars. 

The spectroscopic data were reduced using 
IRAF\footnote{IRAF (Image Reduction and Analysis
Facility) is distributed by the National Optical Astronomy
Observatory, which is operated by the Association of Universities
for Research in Astronomy, Inc., under contract with the National
Science Foundation.} IMRED, CCDRED, and ECHELLE packages to correct
for the bias level, trim the overscan region, extract individual
orders, fit the continuum, apply a wavelength solution using
the Th-Ar spectra (and determine a systematic correction using
the observed radial velocity standard). Master flat field frames
were produced each night, and normalized using APFLATTEN, following
which the data frames were divided by the master flat field frames
prior to extraction of individual orders using APALL. 

The Th-Ar comparison spectra obtained before and after each
program star exposure were used to measure radial velocities.
ECIDENTIFY and DISPCOR were used to identify the lines and
determine the dispersion solution for each order, and CONTINUUM
task enabled us to interactively fit a high-order cubic spline
to produce the continuum-normalized, wavelength-calibrated
spectra. Stars with more than one observed spectrum were 
cross-correlated and then
combined into a single final spectrum using SCOMBINE.
Table~\ref{tab:observations} includes the Heliocentric
Julian Date of mid-exposure as well as the final S/N levels
{\em per pixel} of the combined spectra.

\subsection{Echelle Radial Velocities}

We used Fourier cross-correlation techniques to derive
radial velocities from our echelle spectra, employing
FXCOR within the RV package in IRAF. 
Radial
velocities were determined for many individual orders after
rebinning to common log-linear dispersions, the edges of each
order ($\approx 200$ pixels) were set to zero, and the edges,
$\approx 0.2$ points, were apodized with a cosine bell curve.
The maximum peak of the Fourier transform was fitted by a 
Gaussian, with its center used to determine the radial velocity.
The observed radial velocities were then transformed into
Heliocentric radial velocities using FXCOR. Table~\ref{tab:observations}
includes the final results, the number of orders used, and
the standard error per order.

For the three program stars, the agreement with the results
from our Argus observations is very good, with a mean difference
(echelle $-$ Argus) of +1.3 \kms\ and $\sigma = 1.9$ \kms.

\section{ELEMENTAL ABUNDANCES}

\subsection{Stellar Parameters: Initial Estimates}

It is somewhat challenging to regard a line-rich high-resolution spectrum
and make a good initial guess for the stellar parameters. Photometry
is often invoked for such estimates, but the low Galactic latitude of our
program stars also introduces difficulties. A good initial guess
may be made in the following manner, however. 

We first adopt a reddening estimate based on Schlegel, Finkbeiner,
\& Davis (1998).
The measured radial velocities and the results in Table~\ref{tab:warp}
suggest that the stars lie at large distances, so we adopt
a low metallicity, [Fe/H] $\approx\ -0.6$, based on the results
from Paper~I that found such low metallicities for old
open clusters at
essentially all Galactocentric distances greater than about 12~kpc.
Temperatures and gravities may be estimated using the
color-temperature and color-bolometric correction calibrations of 
Alonso, Arribas, and Martinez Roger (1999). The 2MASS photometry
was transformed to the relevant IR system employed by Alonso et al.\ (1999)
using the equations given by Alonso, Arribas, \& Martinez Roger (1994), 
and Carpenter (2001). We adopted masses
of 1.0 \msun\, and the luminosities were derived using 
distances estimated (naively) from the measured radial velocities.
Initial parameters for all three stars are given
in Table~\ref{tab:parameters}.

\subsection{Refining the Stellar Parameters}

In Paper~I, we demonstrated that our adopted atomic
line parameters and methods of analysis resulted
in good agreement between our derived abundances and
published values for the Sun,
Arcturus, and extensive prior work on the
giants in the old open cluster M67. We refer the reader
to that paper for a more complete discussion of our
procedures, and of the sensitivity of our final
abundances to our adopted values for $T_{\rm eff}$,
log~$g$, and [Fe/H]. In brief, to derive the
stellar parameters, we measured equivalent widths
of Fe~I and Fe~II lines in our spectra. 
Table~\ref{tab:lines} provides the basic data.

We first determined $T_{\rm eff}$ by using the weaker
Fe~I lines so that no trend between the
derived iron abundance and lines' excitation potentials
was present. We then included the stronger Fe~I lines
and adopted a microturbulent velocity, $V_{\rm turb}$, such that
the derived iron abundance was independent of
equivalent width (and hence whether saturation 
of the line was present or not). The gravity was
the final parameter determined, exploiting the
requirement that the pressure-insensitive Fe~I
lines yield the same iron abundance as the
pressure-sensitive Fe~II lines. The final [Fe/H]
abundances are given in Table~\ref{tab:abundances}.

The adopted spectroscopic parameters differ somewhat
from our initial estimates, but considering the
difficulties involved, the agreement is quite good.
The reddening estimates from Schlegel et al.\ (1998)
are high, and hence somewhat uncertain, and of course
they refer to the total along the line of sight,
and over-estimate the true reddening and extinction
for stars closer to the Sun. The initial gravity
estimates are most sensitive to our initial assumptions
about the stars' distances, which we determine on the
basis of measured radial velocities and a radial velocity
vs.\ distance relation that does not account for 
velocity dispersion nor the presence of the Galactic
warp. 

\subsection{Other Elemental Abundances}

We derived abundances for fifteen other elements, using lines
and $gf$ values provided in Table~\ref{tab:lines}. 
For all fifteen
elements, we relied on spectrum synthesis
to derive the elemental abundances for each line. MOOG was employed to
generate 8~\AA\ windows centered on each line and the
elemental abundances were adjusted until the fit was
judged to be optimal. A few elements required special
attention. For Mn, we had to include the effects of
hyperfine splitting, following Prochaska \& McWilliam (2000).
In a number of cases, we also had to consider isotopic
splitting, in which case we adopted solar isotopic abundance
ratios. The elements so considered included Co and Ba
(Prochaska et al.\ 2000), Rb (Lambert \& Luck 1976; Tomkin \& Lambert 1999),
La (Lawler et al.\ 2001a), and Eu (Lawler et al.\ 2001b).

Final abundances are given for our three outer disk field
stars in Table~\ref{tab:abundances}.

\section{DISCUSSION}

\subsection{Distance Estimations}

We are now in a position to substantially improve our distance
estimates for the three program stars. 

Our basic approach is
as follows. First, we estimate the interstellar reddening
to each star by requiring that the de-reddened $V-K$ color
yield the spectroscopically-determined effective temperature
using the color-temperature relations from Alonso et al.\ (1999).
This leads to E($V-K$) and E($B-V$) (= E($V-K$)/2.74).
We are then able to determine $K_{0}$ and we then estimate
$M_{K,0}$, and, hence ($m-M$)$_{0}$, by referring to a cluster
with known reddening and distance and with a similar metallicity
as our program stars.

We have chosen Be~29 as our reference cluster for four
reasons. First, it has a very similar metallicity as our
three field stars, [Fe/H] = $-0.54$ according to Paper~I.
The second advantage is that the cluster's iron abundance
was determined using exactly the same observational and
analysis procedures as employed for the field stars.
Third, its reddening is quite low, estimated to be 
E($B-V$) $\approx\ 0.04$ in Paper~I. Finally, it has
reasonably well-defined optical and infrared color-magnitude
diagrams. We prefer to work with infrared magnitudes because
of their reduced sensitivity to extinction.

Before we undertake the comparisons, however, we apply
the same reddening estimation procedure outlined above
to the cluster. We are able to reproduce the temperatures
and gravities of stars 988 and 673 we derived spectroscopically
in Paper~I if E($B-V$) = 0.16 mag, and for a cluster
distance modulus of 15.76 mag.

With the cluster's de-reddened $K_{0}$ vs.\ $T_{\rm eff}$ diagram,
and our estimated temperatures for the three field stars,
we derive de-reddened $M_{K}$ values, distance moduli,
heliocentric and Galactocentric distances, all of which
are provided in Table~\ref{tab:distances}. For star 9060,
the estimate of $M_{K}$ is especially straightforward since
its temperature and surface gravity indicate it is a
red clump giant, and as discussed in Paper~I, such
stars form the basis for all of our distance estimates
for the clusters. We have adopted the recommendation
of Alves (2000) that $M_{K}$ = $-1.61$ mag for such stars.

\subsection{Comparison with Old Open Clusters}

\subsubsection{Differences in Elemental Abundance Patterns}

Although it appears that the field stars lie at large
Galactocentric distances, their ages are unknown. The first
question we ask is whether the chemical abundances of these 
field stars resemble those of the outer disk old open clusters.
We approach this question first by noting that all three field
stars have very similar abundance patterns, and we therefore
average the element-to-iron abundance ratios for all three stars.
We do the same for the old open clusters studied
in Paper~I, averaging the results
for all six stars, except for the neutron capture elements.
In Paper~I we noted that two of the clusters, Be~31 and NGC~2141,
show significantly elevated abundances of the light $s$-process 
element Zr, the heavy $s$-process elements Ba and La, and 
the $r$-process element Eu.
We therefore have computed three sets of averages
for these elements: all six stars in the four outer disk
clusters, the four stars in Be~20 and Be~29, and the two stars
in Be~31 and NGC~2141. Finally, for comparison purposes, we have
also averaged the results for the three stars we studied in the
old open cluster M67. Our results are summarized
in Table~\ref{tab:comparisons} and Figure~\ref{fig:xfecomparisons}.
We remind the reader that all of these stars were observed 
with similar equipment and analyzed
together, so while systematic errors in the analyses may
compromise comparisons with results of other workers, our
comparisons between the field stars, cluster stars, and
the Sun should be robust.

Perhaps the first result of note is that the old open clusters
and the field stars in the outer disk are essentially identical in their [X/Fe]
abundance distributions. The only significant differences
are in the neutron-capture elements. If we restrict the
comparison of the field stars to the four stars we studied in Be~20
and Be~29, we find that
the field stars are very similar to the cluster stars
in their mean La abundances, but the field stars
have lower Ba abundances and somewhat higher Eu abundances. 

Let us now make comparisons that include the comparable
age but more metal-rich and smaller Galactocentric distance
cluster M67.

The light elements Na and Al are similar in all three sets
of stars: the field stars and the stars in the comparable
age clusters of the outer disk and M67. There are minor and
probably insignificant differences in the iron peak
elements Co and Ni, and Mn may be more deficient relative
to iron in the more metal-poor stars of the outer disk.

The most notable differences between the outer disk field
and cluster stars and those of M67 and the Sun are
in the ``$\alpha$" elements,
O, Mg, Si, Ca, and Ti, and the $r$-process element Eu.
In the field stars, as in the old open clusters of the outer disk,
the [$\alpha$/Fe] and [Eu/Fe] abundance ratios are enhanced.

In summary, the field stars appear to behave very similarly to
the stars in the old open clusters at comparable Galactocentric
distances. This has an interesting consequence regarding the
origins of these field stars. One objection to our finding of
elevated [$\alpha$/Fe] ratios in outer disk stars
is that such behavior is often
seen in stars belonging to the Galactic {\em thick disk}
(Fuhrmann 1998; Prochaska et al.\ 2000; Bensby et al.\ 2003, 2004;
Brewer \& Carney 2004, 2005). Even at the low latitudes at which
we are looking, our three stars have estimated distances from
the Galactic plane of about 300 to 650 pc, far enough to normally qualify
for thick disk membership. However, the thick disk appears to
be uniformly old (Bensby et al.\ 2004; Brewer \& Carney 2004, 2005),
and the ages of the old open clusters are not consistent with
thick disk membership. The very close agreement between the
three field stars studied here and the old open clusters suggests
similar origins, from which we conclude that the field stars
are also unlikely to belong to the thick disk.
 
We now consider what the two sets of stars
reveal about chemical evolution in the outer Galactic disk.

\subsubsection{Trends with Galactocentric Distance}

In Paper~I we drew attention to the apparent disappearance
of the Galactic metallicity gradient for $R_{\rm GC} > 12$ kpc.
The three field stars confirm this behavior, and, in fact, appear
to show that the radial abundance gradient for older
stars and clusters may disappear
at an even smaller Galactocentric distance, perhaps 10~kpc.
This result is hard to understand in simple models of Galactic chemical
evolution. 
A metallicity gradient arises in
the chemical evolution of a closed system, with lower-density
outer regions evolving more slowly than the higher-density
inner regions. At any epoch, the transformation of gas into
stars should be less complete in the outer regions, and
the resultant mean metallicity should be lower. 
But what if the Galaxy is not a closed system?
Twarog, Ashman, \& Anthony-Twarog (1997) addressed this
question in their study of the mean metallicities of a
large sample of open clusters. 
Twarog et al.\ (1997) noticed
that the clusters also showed
a break in the metallicity gradient at $R_{\rm GC} \approx\ 10$~kpc,
with an essentially constant value at larger Galactocentric
distance\footnote{It is interesting that Janes (1979) had
also seen similar behavior in his Figure 10, but, unfortunately, 
his Figure 11 binned the data and erased the change in the metallicity
gradient, supplanting it with an apparently even steeper one!}.
Twarog et al.\ (1997) considered this result and
offered what we believe
to be a viable explanation. 
If the outer disk is the repository of a considerable amount of
gas accreted by the Galaxy's disk over a long period of time,
a metallicity gradient might not be expected.
Indeed, Twarog et al.\ (1997) speculated that the
transition zone, at $R_{\rm GC} \approx\ 10$ kpc, may represent
the boundary of the ``original" Galactic disk. 

Of course, a key question is whether additional data further
support the suggestions of Twarog et al.\ (1997). 
Chen, Hou, \& Wang (2003) suggested that the discontinuity
seen by Twarog et al.\ (1997) is not 
apparent in their study. Their analyses
do show that a linear fit to the
radial metallicity gradient does appear to be steeper for
older clusters. Unfortunately, they do not cite the uncertainties
of the derived metallicity gradients,
nor attempt any non-linear fits that our data indicate
should be undertaken.
Perhaps the best that can be
said at this point, in our opinion, 
is that the plotted data of Chen et al.\ (2003)
are consistent with the results of Twarog et al.\ (1997) and our
own work in Paper~I. 

Our results pose two additional difficulties.
First, our results and those
of Twarog et al.\ (1997) suggest that
there does not appear to be
a significant dispersion in [Fe/H] at
large Galactocentric distances. Not only is there
little to no metallicity gradient, but the three field
stars and three of the four outer disk open clusters we analyzed
in Paper~I
show very similar [Fe/H] values. If we include the three
clusters from Paper~I with $R_{\rm GC} > 12$ kpc,
Be~20 ($R_{\rm GC}$ = 16.0 kpc; [Fe/H] = $-0.45$),
Be~29 ($R_{\rm GC}$ = 22.5 kpc; [Fe/H] = $-0.54$), and
Be~31 ($R_{\rm GC}$ = 12.9 kpc, [Fe/H] = $-0.57$), and the
three field stars, we find $<$[Fe/H]$>$ =
$-0.48$, with a standard error of only 0.07 dex, which is
probably the relative uncertainty of the individual measures.

We can not offer an
explanation for the very similar metallicities, but draw attention to it
as a feature that must be explained along with the
absence of a metallicity gradient. Of course, our
sample is small, and so we might be vulnerable to
the vagaries of statistical chance. However, the
much larger number of photometric metallicity
estimates that form the basis of the results
of Twarog et al.\ (1997) suggest similar behavior.

The other puzzle, and to which we drew
special attention in Paper~I, and which recurs
with the field stars, is the near-uniform enhancement of the
$\alpha$ elements, including oxygen. In Figure~\ref{fig:rgcwarp}
we show this graphically, using the averaged abundances
of Mg, Si, Ca, and Ti with respect to iron, plotted vs.\
Galactocentric distance. The differences with
respect to M67 are modest, somewhat over 0.1 dex, but
relative to the Sun, whose age is comparable to M67 and
the outer disk clusters, the difference is remarkable,
amounting to roughly 0.2 dex. As we discussed in Paper~I, it is extremely
difficult to understand enhancements of these elements,
and of Eu, within closed box models. One expects such enhancements in stars that
formed shortly after star formation commenced, as is seen
in the metal-poor and very old stars of the Galactic halo and thick disk,
because at that point ejecta from Type~II supernovae were
dominating the heavy element enrichment of the interstellar
medium. Solar [$\alpha$/Fe] ratios are achieved at a much
later time, once Type Ia supernovae have had sufficient
time (perhaps one to a few billion years) to contribute
ejecta more enriched in iron than in the $\alpha$ or
$r$-process elements. The problem for the clusters,
and, we assume, for the field stars in the outer disk is
that the old open clusters are not as old as the halo
or metal-poor stars of the thick disk, but have
ages of less than 6~Gyrs, comparable to that of the Sun. 
If SNe~II have played a major
role in the nucleosynthetic history of the outer disk,
they have done so relatively recently, and not long
before these clusters (and field stars) formed. SNe~Ia
have contributed as well, since the [$\alpha$/Fe] 
and [Eu/Fe] are not as high as are found in the
Galactic halo, and in Paper~I we noted that AGB stars
have also contributed to the interstellar medium, judging
by the significant abundances of elements largely produced
by $s$-process nucleosynthesis. (The same must be true
for our field stars since they share similar abundance
patterns as the cluster stars.) One way to explain
these observations is to assume that the growth of the
Galactic disk
has been intermittent, and that merger events, be they
large or small, have triggered episodes of star
formation so that SNe~II have been able to contribute
anew to the chemistry of the interstellar medium, unlike
the case in the Solar neighborhood, where [$\alpha$/Fe]
and [Eu/Fe] are, essentially, solar. The star formation
may have continued long enough to enable the
interstellar medium to have become enriched by AGB stars
and also Type Ia supernovae, but not to the same degree
as are seen in the thin disk stars of the solar neighborhood.
We look forward to chemical evolution models that may help
explain our observations.

Let us consider the ``growth of the Galactic disk" and merger
models a bit further.
In Paper~I we compared the velocities of the old open clusters
with the ``GASS" (Global Anticenter Stellar Structure) discussed
by Frinchaboy et al.\ (2004), which may be related to the Monoceros
Ring (Newburg et al.\ 2002; Ibata et al.\ 2003; Rocha-Pinto et al.\ 2003;
Yanny et al.\ 2003) and the possible ``Canis Major galaxy" (Martin 
et al.\ 2004). Table~\ref{tab:observations} includes the
``Galactic System Reference" velocity, $V_{\rm GSR}$. Comparison
with Figure~2 of Frinchaboy et al.\ (2004) indicates that all three field
stars have velocities consistent with membership in the
putative GASS. We remind the reader
that GASS does not differ markedly in inclination from the Galactic disk
($\approx 17$ degrees according to Frinchaboy et al.\ 2004;
$25 \pm 5$ degrees, according to Pe\~{n}arrubia et al.\ 2004)
and since both respond to the same Galactic gravitational potential,
radial velocities may not provide a particularly good discriminant
regarding membership. 

If the old open clusters in the outer disk or the three
field stars studied here are to be associated with the GASS,
there is a curious difference in the mean metallicities and
metallicity spreads. Yanny et al.\ (2003) suggested the
the mean [Fe/H] value for the stars in the Monoceros Ring,
which is thought to be related to the GASS, is near $-1.6$,
much more metal-poor than we have found in the outer disk stars
and clusters. The study of field M giants to identify the
Monoceros Ring by Rocha-Pinto et al.\ (2003) 
and Crane et al.\ (2003) suggests that metal-rich stars may
also be part of the GASS, and therefore that
the metallicity range in the GASS may be very large. This is
inconsistent with what we have found among
our limited study of outer disk clusters
and stars. Of course, the simplest explanation is that 
the objects we have studied do not belong to the GASS, but
are representative of the Galactic outer disk.

The similar [Fe/H] and [X/Fe] patterns seen
among the three field stars indicates similar chemical enrichment
histories, but we note again that the enhancements of [$\alpha$/Fe]
are identical in Be~29 (a GASS member) and Be~31 (a non-member).
Further, the globular cluster NGC~2298 has been
suggested to be a member of GASS, and it
shows similar elevated [$\alpha$/Fe] values despite being several
Gyrs older than any of the open clusters we have studied. We
have invoked the idea of episodic star formation to explain
enhancements of [$\alpha$/Fe] seen in the younger open clusters,
but do we expect this to happen in small galaxies which are
ultimately captured by the Milky Way? Or is it more likely
that the triggered star formation happens when the smaller
galaxy finally merges with the Milky Way? We believe the
second option is the more viable explanation. As Venn et al.\ (2004)
have discussed, the current dwarf galaxies surrounding the
Milky Way show solar (and even sub-solar) [$\alpha$/Fe]
abundance ratios and at very low metallicities. It would
be surprising if the recently merged or merging dwarf
galaxy that contains NGC~2298 and a host of younger and
more metal-rich open clusters would deviate from this trend.

An attractive alternate view,
in our opinion, is that the outer disk has experienced a number
of smaller accretion events, involving metal-poor gas, and
which triggered bursts of star formation so that
SNe~II helped rapidly enrich the interstellar medium and
produce the enhanced abundances of $\alpha$ and $r$-process
elements.
If the Canis Major galaxy is a real structure (and our three
field stars lie very near the direction to that galaxy's
center, according to Martin et al.\ 2004), this may only
be the latest such accretion event. In this view, the
Galaxy's disk has grown steadily with time, as
Twarog et al.\ (1997) described. 

One way to explore this further, perhaps, is to consider stars that have
formed more recently, out of the more recently enlarged
Galactic disk. 

\subsection{Comparison with Young Stars \& Nebulae: Cepheids}

Paper~I concentrated on the older
known clusters with large Galactocentric distances, and in this
paper we have considered field stars with unknown but probably large, ages. 

We concentrate
on the extensive and thorough abundance analyses of Cepheid
variable stars undertaken
by Andrievsky et al.\ (2002a,b,c; 2003) and Luck et al.\ (2003).
In Figure~\ref{fig:cepheidswarp} we repeat the results for older
open clusters in the outer Galactic disk, the three field stars
from this paper, and all the Cepheids studied, except that 
we eliminated
redundant listings of stars, and
retained only those stars for which the abundances
of Mg, Si, Ca, and Ti are available from these papers. 

We focus first on the behavior of [Fe/H] vs.\ $R_{\rm GC}$.
There are three notable points in Figure~\ref{fig:cepheidswarp}. First, 
the Cepheids appear to show a metallicity gradient, with the
lowest value reached for EE~Mon at $R_{\rm GC} \approx\ 15.1$ kpc.
Second, the gradient may not be as simple as a linear
decline in [Fe/H] with $R_{\rm GC}$. One may argue that beyond
about 12 kpc, there is no real gradient, if one ignores EE~Mon.
Andrievsky et al.\ (2002c), Luck et al.\ (2003), and
Andrievsky et al.\ (2003) argued that the change in [Fe/H] trends (or lack
thereof) beyond
$R_{\rm GC} \approx\ 10$ kpc may be due the effects of
the Galaxy's co-rotation point, where the spiral arm pattern
speed matches the stellar circular speeds. We argued in Paper~I
that while this model has some merit, it does not seem capable of 
explaining the same lack of a metallicity gradient in the old
open clusters (and now seen as well in the three field stars)
because the behavior extends over 
much larger Galactocentric distances, from 10 to over 20 kpc.
Let us consider, then, that the metallicity gradient seen in
the Cepheids is real, and that it reaches a fairly low
metallicity, apparently [Fe/H] $\approx\ -0.2$, 
at $R_{\rm GC} \approx\ 12$ kpc. (It may yet drop again, based on
the most distant star, EE~Mon.)
If this version of the Cepheid radial metallicity gradient is correct, it implies
that the younger stars show the same behavior as the older
stars, but the metallicity trend is displaced to larger Galactocentric distances,
and perhaps to high metallicities. 
This is consistent with the idea that the disk has grown
radially over time, and that its mean metallicity has risen
as well (but, again, recall the result for EE~Mon).

Figure~\ref{fig:cepheidswarp} also shows the trend in the
alpha elements' abundances relative to iron, as a function of $R_{\rm GC}$. The
bulk of the cepheids (again excluding EE~Mon) show an
essentially constant value for [$\alpha$/H], consistent with no
metallicity gradient. Since comparisons of metallicity
gradients often compare results for iron from studies
of clusters with abundances of the lighter $\alpha$ elements from abundances of H~II
regions, it is worth keeping in mind the different nucleosynthetic
origins of these elements, and making direct comparisons of
elements created in the same processes.

We turn now to the question of enhanced abundances of
elements manufactured in SNe~II, relative to
those produced in SNe~Ia, in particular [$\alpha$/Fe]
vs.\ $R_{\rm GC}$.
Figure~\ref{fig:cepheidswarp} shows that 
[$\alpha$/Fe] in the Cepheids increases at larger
Galactocentric distances, beyond about 12~kpc, and that this rise is similar
in slope to that seen in the older stars and clusters,
except that the rise begins at a somewhat larger
Galactocentric distance in the younger Cepheids than in
the older clusters and field stars.
This is also
consistent with the idea
that the disk is continuing to grow, but that the
effects of star formation triggered by infall are most
pronounced at Galactocentric distances of 15 kpc (and perhaps
beyond?) rather than at 10 kpc as was the case in the past.
Again, this is qualitatively consistent with a growing
Galactic disk. (And it is hard to explain in terms
of the co-rotation point model of Luck et al.\ 2003
and Andrievsky et al.\ 2003.) We need careful (and consistent!)
abundance analyses of the few Cepheids with $R_{\rm GC}$ values
of 15 kpc and larger to explore these trends further.

In a future paper we will present results for a small sample of
distant but very young and easily studied Cepheid variables
of the outer disk. It is, nonetheless, worthwhile to consider
what is available already in the literature. 

\subsection{Comparison with Young Stars \& Nebulae: B Stars \& H~II Regions}

Cepheids have several advantages, being cool enough to provide absorption
lines for many elements, and such stars also have well-determined distances.
Cepheids have one problem, however, in that mixing may alter the
abundances of some key elements, such as carbon, nitrogen, and oxygen.
It is therefore worth exploring the Galactic disk's metallicity gradient
using other tracers of recent star formation.

Shaver et al.\ (1983) provided a comprehensive study of Galactic H~II
regions whose $R_{\rm GC}$ values ranged from about 4 to almost 14 kpc,
and Fich \& Silkey (1991) extended the work to roughly 18 kpc. While
the scatter for any one measurement is relatively large, the data
are consistent with what is seen for the Cepheids: the oxygen abundances
reach a basement value (as do those of sulfur and nitrogen) at
between 10 and 12 kpc. More recent work by V\'{i}lchez \& Esteban (1996)
shows that the five H~II regions they analyzed with $R_{\rm GC} > 15$~kpc
have very similar oxygen abundances, consistent with the
trend seen for the $\alpha$ elements seen by Fich \& Silkey (1991) and in
Figure~\ref{fig:cepheidswarp}. Henry \& Worthey (1999)
have provided a good summary of all published work, and these
trends are confirmed, although there is considerable scatter
in the oxygen abundances at any Galactocentric distance.
It is unclear if the scatter seen is due to true changes in
metallicity or systematics of the analyses. The
approximate constant, but lower, abundances of oxygen seen in H~II regions
beyond 15~kpc or so is intriguing, and it is therefore
even more important to push the study of [Fe/H] and
[$\alpha$/Fe] to such large distances using such tracers.

Rolleston et al.\ (2000) studied abundances in 80 B~stars
in 19 young open clusters, about half of which have 
$R_{GC} > 10$~kpc. They considered the suggestion by
Twarog et al.\ (1997) that a step function exists in
the Galactic metallicity gradient. They found that 
magnesium abundances appear
to be independent of $R_{\rm GC}$, at least for distances
outside 10 to 12 kpc, which is consistent with 
Figure~\ref{fig:cepheidswarp}. However, two of the other
$\alpha$ elements, oxygen and silicon, do not show as
clear a trend, but considerable scatter at distances
beyond 12 kpc.

More recently, Daflon \& Cunha (2004) have
provided a very careful exploration of all available Galactic B star data,
seeking signs of differences as a function of $T_{\rm eff}$ of
the program stars or their rotational velocities, finding no significant
effects among the stars they studied. Further, most of the
elements studied behave similarly, which means that we may explore
a mean $\alpha$ element abundance 
as a function of $R_{\rm GC}$. (Unfortunately, iron abundances
are not available for these stars.)
We adopt their abundances for
oxygen, magnesium, silicon, and sulfur, and then correct them
relative to solar photospheric abundances taken from
Grevesse \& Sauval (1998). We take straight unweighted means
of these four elemental abundances, discarding any star for which
the abundances for the elements is incomplete. Figure~\ref{fig:bstars}
summarizes the inner and outer disk abundances. While it is certainly
possible to draw a straight line fit, we assert that the data
are also consistent, within the uncertainties, with the
results of the Cepheids seen in Figure~\ref{fig:cepheidswarp}.

\subsection{Comparison with Other Galaxies}

Is our Galaxy special or do similar transitions between
well-defined metallicity gradients and their apparent
disappearance at some transition radius occur in other
disk galaxies? An exhaustive review of extra-galactic metallicity
gradients is inappropriate in this paper, but we draw
attention to some relevant points.

Metallicity gradients have routinely been discussed in
linear terms, either changes in logarithmic abundance 
vs.\ radial distance (Pagel \& Edmunds 1981) or
fractional isophotal diameter (Zaritsky, Kennicutt, \&
Huchra 1994). Our point, that linear trends may not
be applicable to the total extent of our 
Galaxy's disk, appears to also be
the case in at least some other galaxies. Zaritsky
et al.\ (1994), in particular, questioned whether
exponential behavior might be more descriptive for some
galaxies. We speculate that if accretion is the
cause for the lack of a gradient in our Galaxy's
outer disk, it would be worth studying galaxies showing
similar behavior to ascertain if such a characteristic
is related to some other property, such as cluster
environment, or, possibly, galactic mass. We suggest
that galaxies in denser environments or with larger
masses will have more opportunities for accretion
events to alter the metallicity gradients in the
outer disks.

Metallicity gradients alone will not provide the
insight we seek, however. The nucleosynthetic
history is revealed in element-to-iron ratios,
and iron abundances are rarely obtained in the studies of H~II
regions, which provide the bulk of the results
in the studies of other galaxies. We choose
to concentrate on the most obvious comparison
galaxy, M31 (= NGC~224). Here, at least, we have
more detailed elemental abundances available
for the H~II regions and also a limited amount
of information from abundance analyses of
individual stars, including A supergiants (Venn et al.\ 2000)
and B supergiants (Trundle et al.\ 2002). Like
H~II regions, these stars represent a snapshot of
the recent star formation history in the disk
of M31. Abundances of $\alpha$ elements
in older disk stars, represented by normal K giants,
are not available, and abundances of planetary nebulae
outside the bulge of M31 (see Jacoby \& Ciardullo 1999)
are very few in number. 

We begin by considering the metallicity gradient, represented
by the element oxygen. Figure~\ref{fig:m31otoh} shows the
logarithmic abundances of oxygen from studies of H~II
regions (Dennefeld \& Kunth 1981; Blair, Kirshner, \& Chevalier 1982)
and the stars studied by Venn et al.\ (2000) and Trundle
et al.\ (2002). We again transform logarithmic abundances relative
to hydrogen to values relative to the Sun using the solar
abundances from Grevesse \& Sauval (1998).
Taking all 17 H~II regions, there is a detectable
metallicity gradient, 
$\Delta [O/H] = -0.028 \pm 0.007$ dex kpc$^{-1}$.
Adding in the ten stars yields
$\Delta [O/H] = -0.017 \pm 0.008$ dex kpc$^{-1}$. But M31's
young disk population, like the old population of the Milky Way, appears to
show a shallower, or even absent, metallicity gradient beyond
a distance of about 10~kpc. The 14 H~II regions lead to
$\Delta [O/H] = -0.012 \pm 0.010$ dex kpc$^{-1}$, while
the total sample of 17 H~II regions and stars yields
$\Delta [O/H] = -0.006 \pm 0.008$ dex kpc$^{-1}$.

The stellar abundances provide one more interesting piece of
evidence. Figure~\ref{fig:m31ofe} presents the very limited
data regarding the relative contributions of SNe~II, represented
by oxygen, relative to the contributions of SNe~Ia, represented
by iron. A single datum with enhanced [O/Fe] at large
galactocentric distance hardly proves the similarity of
M31 to the Galaxy, but at least it does point the way to
future work.

\section{CONCLUSIONS}

We have used color-magnitude diagrams and radial velocities obtained
from multi-fiber spectroscopy to identify candidate red giants of
the outer Galactic disk. High-resolution, high-S/N echelle
spectrscopy has confirmed these identifications for three stars
lying in the direction of the southern Galactic warp. The [Fe/H]
values and Galactocentric distances of these stars confirms the
results found in Paper~I. At Galactocentric distances beyond 10 to 12 kpc,
the Galactic metallicity gradient for the moderately old
stars and clusters disappears. Further, abundances of
elements manufactured primarily in SNe~II are enhanced in both
the field stars and the old open clusters of the outer Galactic disk,
despite the fact that none of the clusters have ages greater
than about 6~Gyrs.
In the solar neighborhood, abundances of the elements produced largely
in SNe~II are much lower in stars and clusters with
these ages, showing the outer disk has experienced
a very different star formation history. We suggest that the difference
is caused by a gradual growth of the outer disk. Episodic accretion
can erase a metallicity gradient, and if the accretion events trigger
bursts of star formation over a period of time, it may also
explain the elevated abundances of the $\alpha$ elements and the
$r$-process element Eu. The available data for the much younger
Cepheids suggest similar trends may be at work there, too, but,
in essence, displaced to larger Galactocentric distances. H~II
regions and B~stars appear to support the idea of a reduced or
absent metallicity gradient among young objects in the outer disk.
This, too,
is consistent with the steady growth and episodic chemical enrichment
of the outer disk. The limited data for young stars and H~II regions in
the disk of M31 hint at similar behavior.

Future measurements of iron and $\alpha$ element abundances are
certainly desirable for more stars across the disk of M31, as
well as for M33, and for
both younger and older stars in the outer disk of the Milky Way.
Cepheid variables with $R_{\rm GC} \geq\ 15$~kpc would be
most welcome, as would abundance of objects more clearly associated
with the Monoceros Ring and other star streams. 

\acknowledgments

We are extremely grateful to the National Science Foundation for
their financial support through grants AST-9619381,
AST-9988156, and AST-0305431 to the University of North
Carolina.

\begin{deluxetable}{rrrr}
\tablenum{1}
\tablecolumns{4}
\tablecaption{Predicted radial velocities relative to the Local
Standard of Rest for different Galactocentric distances toward
$\ell = 245.75$, $b = -4.1$. \label{tab:warp}}
\tablewidth{0pt}
\tablehead{
  \colhead{d} &
  \colhead{$R_{\rm GC}$} & 
  \colhead{$V_{\rm LSR}$} & 
  \colhead{$V_{\rm rad}$} \\
  \colhead{kpc} &
  \colhead{kpc} &
  \colhead{\kms} &
  \colhead{\kms}
}
\startdata
 5.0 & 11.0 & 55 & 73 \\
 7.5 & 13.0 & 77 & 96 \\
10.0 & 15.2 & 95 & 113 \\
12.5 & 17.4 & 108 & 126 \\
15.0 & 19.7 & 119 & 137 \\
17.5 & 22.0 & 127 & 146 \\
20.0 & 24.4 & 135 & 153 \\
\enddata
\end{deluxetable}

\clearpage

\begin{deluxetable}{lrrrrrr}
\tabletypesize{\footnotesize}
\tablenum{2}
\tablecolumns{7}
\tablecaption{Photometry of field red giants in the direction of the 
galactic southern warp\label{tab:warpphot}}
\tablewidth{0pt}
\tablehead{
   \colhead{Star ID} & \colhead{x} & 
   \colhead{y} & \colhead{$V$} &  \colhead{$\sigma$} &
   \colhead{$V-I_{C}$} &
   \colhead{$\sigma$}
}
\startdata
  6194 & 632.35 & 549.11 & 12.069 & 0.001 & 0.852 & 0.001 \\
  7245 & 743.09 & 639.93 & 12.238 & 0.001 & 0.654 & 0.001 \\
  8235 & 568.60 & 723.38 & 12.263 & 0.001 & 0.839 & 0.001 \\
  8414 & 821.97 & 739.55 & 12.294 & 0.001 & 0.618 & 0.001 \\
  9037 & 970.74 & 791.66 & 12.518 & 0.001 & 0.643 & 0.001 \\
  8621 & 909.56 & 758.48 & 12.570 & 0.001 & 0.524 & 0.001 \\
  6326 & 729.49 & 559.45 & 12.583 & 0.001 & 0.852 & 0.001 \\
  8622 & 954.65 & 757.60 & 12.647 & 0.001 & 0.641 & 0.001 \\
  7013 & 955.57 & 617.85 & 12.698 & 0.001 & 0.717 & 0.001 \\
  6403 & 704.46 & 566.64 & 12.725 & 0.001 & 0.804 & 0.001 \\
  6266 & 945.53 & 554.63 & 12.820 & 0.001 & 1.224 & 0.001 \\
  4897 & 933.43 & 435.79 & 12.901 & 0.002 & 0.574 & 0.003 \\
  6380 & 781.51 & 564.95 & 12.913 & 0.002 & 1.065 & 0.002 \\
  6688 & 874.70 & 590.07 & 12.915 & 0.002 & 1.699 & 0.002 \\
  1178 & 977.60 & 110.65 & 12.928 & 0.002 & 0.683 & 0.003 \\
 10289 & 605.47 & 899.63 & 12.934 & 0.002 & 1.119 & 0.002 \\
  6216 & 847.26 & 550.49 & 12.986 & 0.002 & 0.590 & 0.003 \\
  4021 & 698.82 & 362.46 & 13.033 & 0.002 & 0.657 & 0.003 \\
  6062 & 933.35 & 536.45 & 13.045 & 0.002 & 0.937 & 0.003 \\
  8766 & 594.12 & 771.19 & 13.112 & 0.002 & 0.577 & 0.003 \\
\enddata
\tablecomments{Table \ref{tab:warpphot} is published in its entirety in 
the electronic edition of the {\it Astrophysical Journal}. A portion 
is shown here for guidance
regarding its form and content.}
\end{deluxetable}

\clearpage

\begin{deluxetable}{lrrrrrrrrrl}
\tabletypesize{\footnotesize}
\tablenum{3}
\tablecolumns{11}
\tablecaption{Radial Velocities and Optical/IR
Photometry for Stars Observed with the Argus Spectrograph
\label{tab:argusvels}}
\tablewidth{0pt}
\tablehead{
   \colhead{Star ID} & 
   \colhead{R.\ A.} & \colhead{DEC.} & \colhead{$V$} & 
   \colhead{$V-I_{C}$} & \colhead{$K$\tablenotemark{a}} & 
   \colhead{$J-K$\tablenotemark{a}} & \colhead{$H-K$\tablenotemark{a}} &
   \colhead{$V-K$} & \colhead{$V_{\rm rad}$} &
   \colhead{Notes}
}
\startdata
11015 &  7:39:35.1 & $-31$:08:31 & 13.48 &  1.32 & 12.81 &  0.21 &  0.03 &  0.67 &   69.3 &  1 \\
 9999 &  7:39:46.9 & $-30$:59:12 & 14.00 &  1.33 & 10.86 &  0.70 &  0.14 &  3.14 &  112.7 &  \\
 5234 &  7:40:46.1 & $-30$:43:45 & 14.06 &  1.49 & 10.40 &  0.89 &  0.18 &  3.66 &   93.7 &  \\
 4338 &  7:40:55.5 & $-31$:08:54 & 14.06 &  1.43 & 10.64 &  0.81 &  0.13 &  3.42 &   85.5 &  \\
 5218 &  7:40:45.9 & $-30$:51:15 & 14.06 &  1.30 & 10.92 &  0.72 &  0.10 &  3.14 &   96.2 &  \\
 7647 &  7:40:15.6 & $-30$:55:11 & 14.06 &  1.25 & 11.15 &  0.68 &  0.11 &  2.91 &   70.8 &  \\
 6884 &  7:40:25.8 & $-30$:44:21 & 14.10 &  1.41 & 10.74 &  0.80 &  0.15 &  3.36 &   68.4 &  \\
 7904 &  7:40:12.8 & $-30$:51:22 & 14.10 &  1.32 & 10.91 &  0.74 &  0.15 &  3.19 &   67.9 &  \\
 4199 &  7:40:57.6 & $-31$:01:49 & 14.16 &  1.01 & 11.83 &  0.56 &  0.08 &  2.33 &   17.6 &  \\
 9830 &  7:39:48.8 & $-31$:01:29 & 14.22 &  1.32 & 10.99 &  0.80 &  0.21 &  3.23 &   62.9 &  \\
 7116 &  7:40:22.9 & $-30$:40:27 & 14.25 &  1.32 & 11.13 &  0.71 &  0.13 &  3.12 &   73.8 &  \\
 7559 &  7:40:17.6 & $-30$:40:07 & 14.28 &  1.45 & 10.81 &  0.84 &  0.18 &  3.47 &  114.3 &  \\
  879 &  7:41:40.2 & $-30$:48:32 & 14.29 &  1.70 & 10.18 &  1.00 &  0.21 &  4.11 &   96.0 &  \\
 3917 &  7:41:00.9 & $-31$:04:01 & 14.30 &  1.22 & 11.43 &  0.65 &  0.15 &  2.87 &   75.5 &  \\
 3373 &  7:41:07.5 & $-31$:03:41 & 14.32 &  1.29 & 11.25 &  0.75 &  0.14 &  3.07 &  100.0 &  \\
11027 &  7:39:35.2 & $-30$:50:29 & 14.36 &  1.37 & 11.04 &  0.79 &  0.18 &  3.32 &   83.9 &  \\
  868 &  7:41:40.3 & $-30$:48:46 & 14.38 &  1.42 & 11.02 &  0.77 &  0.22 &  3.36 &  107.9 &  \\
 9878 &  7:39:48.4 & $-30$:55:43 & 14.39 &  1.64 & 10.38 &  1.03 &  0.22 &  4.01 &   64.6 &  \\
 5181 &  7:40:46.8 & $-30$:40:27 & 14.40 &  1.27 & 11.35 &  0.73 &  0.15 &  3.05 &   93.7 &  \\
 6689 &  7:40:28.4 & $-30$:41:28 & 14.42 &  1.47 & 10.90 &  0.83 &  0.16 &  3.52 &   76.6 &  \\
 1543 &  7:41:31.7 & $-30$:48:34 & 14.45 &  1.70 & 10.36 &  1.00 &  0.19 &  4.09 &  112.7 &  \\
 8744 &  7:40:02.0 & $-30$:54:54 & 14.46 &  1.35 & 11.12 &  0.80 &  0.18 &  3.34 &  100.8 &  \\
 3782 &  7:41:02.8 & $-30$:56:08 & 14.47 &  1.18 & 11.62 &  0.68 &  0.14 &  2.85 &   87.5 &  \\
 5052 &  7:40:46.9 & $-31$:08:13 & 14.49 &  1.59 & 10.71 &  0.91 &  0.20 &  3.78 &   85.5 &  \\
 4822 &  7:40:50.1 & $-31$:01:09 & 14.50 &  1.39 & 11.16 &  0.82 &  0.16 &  3.34 &   43.3 &  \\
 2241 &  7:41:22.1 & $-30$:54:08 & 14.50 &  1.27 & 11.55 &  0.67 &  0.15 &  2.95 &   96.8 &  \\
11350 &  7:39:31.4 & $-30$:53:29 & 14.51 &  1.61 & 10.54 &  1.00 &  0.16 &  3.97 &   78.8 &  \\
 2854 &  7:41:14.9 & $-30$:47:55 & 14.52 &  1.42 & 11.15 &  0.80 &  0.12 &  3.37 &  101.3 &  \\
10635 &  7:39:39.8 & $-30$:51:09 & 14.52 &  1.42 & 11.03 &  0.90 &  0.16 &  3.49 &   75.7 &  \\
 6673 &  7:40:27.4 & $-31$:01:53 & 14.53 &  1.36 & 11.23 &  0.83 &  0.19 &  3.30 &   40.5 &  \\
 9060 &  7:39:58.3 & $-30$:57:40 & 14.54 &  1.23 & 11.60 &  0.65 &  0.11 &  2.94 &  130.4 &  \\
 8368 &  7:40:07.1 & $-30$:48:54 & 14.56 &  1.34 & 11.41 &  0.67 &  0.14 &  3.15 &   59.4 &  \\
10330 &  7:39:42.8 & $-30$:56:39 & 14.59 &  1.29 & 11.46 &  0.75 &  0.19 &  3.13 &   93.0 &  \\
 7530 &  7:40:16.5 & $-31$:05:30 & 14.59 &  1.27 & 11.54 &  0.74 &  0.16 &  3.05 &   75.7 &  \\
  263 &  7:41:47.7 & $-30$:59:42 & 14.60 &  1.13 & 11.98 &  0.66 &  0.09 &  2.62 &   41.5 &  \\
 3697 &  7:41:04.8 & $-30$:39:24 & 14.62 &  1.27 & 11.62 &  0.65 &  0.13 &  3.00 &   57.9 &  \\
 6554 &  7:40:28.7 & $-31$:01:12 & 14.62 &  1.36 & 11.38 &  0.76 &  0.13 &  3.24 &   54.6 &  \\
 6755 &  7:40:27.5 & $-30$:41:44 & 14.63 &  1.28 & 11.60 &  0.71 &  0.16 &  3.03 &   50.5 &  \\
11102 &  7:39:34.7 & $-30$:45:47 & 14.65 &  1.42 & 11.10 &  0.89 &  0.20 &  3.55 &   92.6 &  \\
10808 &  7:39:37.5 & $-30$:56:02 & 14.69 &  1.26 & 11.67 &  0.69 &  0.17 &  3.02 &   75.3 &  \\
 5153 &  7:40:46.5 & $-30$:53:32 & 14.70 &  1.23 & 11.76 &  0.69 &  0.16 &  2.94 &   68.9 &  \\
 5814 &  7:40:37.5 & $-31$:08:04 & 14.71 &  1.18 & 11.91 &  0.64 &  0.10 &  2.80 &   59.3 &  \\
 7333 &  7:40:20.2 & $-30$:42:11 & 14.72 &  1.70 & 10.64 &  0.98 &  0.18 &  4.08 &   40.3 &  \\
 1336 &  7:41:34.4 & $-30$:43:40 & 14.72 &  1.55 & 10.99 &  0.90 &  0.23 &  3.73 &   87.1 &  \\
 8216 &  7:40:09.7 & $-30$:42:28 & 14.73 &  1.36 & 11.54 &  0.74 &  0.13 &  3.19 &   33.0 &  \\
 1164 &  7:41:35.4 & $-31$:07:28 & 14.73 &  1.32 & 11.66 &  0.68 &  0.10 &  3.07 &   53.9 &  \\
10044 &  7:39:46.3 & $-30$:58:37 & 14.74 &  1.30 & 11.62 &  0.75 &  0.19 &  3.12 &   93.9 &  \\
 2831 &  7:41:15.2 & $-30$:47:20 & 14.74 &  1.38 & 11.51 &  0.79 &  0.15 &  3.23 &  109.9 &  \\
 4735 &  7:40:51.0 & $-31$:05:11 & 14.75 &  1.26 & 11.78 &  0.71 &  0.14 &  2.97 &   41.7 &  \\
 1465 &  7:41:31.7 & $-31$:08:40 & 14.75 &  1.36 & 11.53 &  0.81 &  0.17 &  3.22 &  101.3 &  \\
 1507 &  7:41:32.0 & $-30$:53:09 & 14.78 &  1.10 & 12.16 &  0.68 &  0.10 &  2.62 &  111.9 &  \\
 2547 &  7:41:17.6 & $-31$:08:08 & 14.79 &  1.07 & 12.45 &  0.45 &  0.13 &  2.34 &   44.0 &  \\
 8516 &  7:40:04.8 & $-30$:55:10 & 14.80 &  1.34 & 11.67 &  0.72 &  0.10 &  3.13 &   91.6 &  \\
 7073 &  7:40:23.1 & $-30$:47:50 & 14.81 &  1.04 & 12.19 &  0.66 &  0.14 &  2.62 &  133.1 &  \\
  384 &  7:41:46.1 & $-31$:03:48 & 14.83 &  1.41 & 11.53 &  0.77 &  0.17 &  3.30 &   95.8 &  \\
 5022 &  7:40:47.3 & $-31$:08:57 & 14.84 &  1.36 & 11.60 &  0.77 &  0.14 &  3.23 &   97.0 &  \\
 7560 &  7:40:17.5 & $-30$:39:19 & 14.84 &  1.81 & 10.55 &  1.03 &  0.18 &  4.29 &   67.2 &  \\
 1532 &  7:41:32.0 & $-30$:41:04 & 14.85 &  1.37 & 11.65 &  0.69 &  0.12 &  3.20 &   73.4 &  \\
 8855 &  7:40:00.7 & $-30$:56:01 & 14.85 &  1.35 & 11.77 &  0.71 &  0.15 &  3.08 &   88.0 &  \\
10195 &  7:39:44.7 & $-30$:52:51 & 14.85 &  1.34 & 11.60 &  0.80 &  0.20 &  3.25 &   94.9 &  \\
 8910 &  7:39:59.5 & $-31$:05:52 & 14.85 &  1.21 & 12.00 &  0.74 &  0.18 &  2.85 &   42.9 &  \\
 3470 &  7:41:06.1 & $-31$:05:58 & 14.91 &  1.14 & 12.19 &  0.70 &  0.12 &  2.72 &   60.2 &  \\
 6789 &  7:40:26.6 & $-30$:53:11 & 14.91 &  1.16 & 12.22 &  0.62 &  0.09 &  2.69 &   55.4 &  \\
10610 &  7:39:39.8 & $-30$:55:21 & 14.91 &  1.32 & 11.69 &  0.76 &  0.16 &  3.22 &   93.6 &  \\
 5034 &  7:40:47.8 & $-30$:54:02 & 14.92 &  1.38 & 11.59 &  0.82 &  0.11 &  3.33 &   75.3 &  \\
 8403 &  7:40:07.2 & $-30$:40:00 & 14.92 &  1.46 & 11.52 &  0.78 &  0.16 &  3.40 &   76.4 &  \\
 9305 &  7:39:55.8 & $-30$:47:21 & 14.93 &  1.02 & 12.63 &  0.48 &  0.09 &  2.30 &   27.5 &  \\
 2025 &  7:41:25.1 & $-30$:53:11 & 14.93 &  1.29 & 11.90 &  0.68 &  0.08 &  3.03 &   94.5 &  \\
  446 &  7:41:46.0 & $-30$:45:57 & 14.94 &  1.67 & 11.10 &  0.85 &  0.21 &  3.84 &  118.7 &  \\
 6756 &  7:40:27.7 & $-30$:39:57 & 14.94 &  1.27 & 11.97 &  0.83 &  0.17 &  2.97 &   83.3 &  \\
 3234 &  7:41:10.0 & $-30$:46:34 & 14.95 &  1.60 & 11.14 &  0.90 &  0.21 &  3.81 &   93.5 &  \\
 3278 &  7:41:09.4 & $-30$:49:18 & 14.95 &  1.70 & 10.93 &  1.01 &  0.26 &  4.02 &  119.2 &  \\
 6447 &  7:40:29.7 & $-31$:04:54 & 14.97 &  1.42 & 11.52 &  0.84 &  0.18 &  3.45 &   75.6 &  \\
 1045 &  7:41:37.5 & $-31$:00:41 & 15.02 &  1.08 & 12.53 &  0.61 &  0.06 &  2.49 &   59.2 &  \\
10148 &  7:39:45.9 & $-31$:08:48 & 15.02 &  1.31 & 12.23 &  0.74 &  0.17 &  2.79 &   56.4 &  \\
  353 &  7:41:47.1 & $-30$:51:38 & 15.02 &  1.42 & 11.73 &  0.78 &  0.16 &  3.29 &   44.0 &  \\
  283 &  7:41:48.0 & $-30$:49:21 & 15.06 &  1.79 & 10.77 &  1.07 &  0.26 &  4.29 &  129.6 &  \\
 7512 &  7:40:17.9 & $-30$:43:45 & 15.07 &  1.55 & 11.34 &  0.93 &  0.19 &  3.73 &  140.6 &  \\
 4884 &  7:40:49.9 & $-30$:52:30 & 15.07 &  1.35 & 11.88 &  0.71 &  0.09 &  3.19 &  114.7 &  \\
 8181 &  7:40:10.0 & $-30$:44:13 & 15.07 &  1.37 & 12.10 &  0.71 &  0.17 &  2.97 &   63.7 &  \\
 7632 &  7:40:16.2 & $-30$:46:13 & 15.10 &  1.10 & 12.80 &  0.50 &  0.17 &  2.30 &   60.6 &  \\
  806 &  7:41:40.5 & $-31$:00:35 & 15.10 &  1.64 & 11.26 &  0.89 &  0.19 &  3.84 &  104.2 &  \\
 1261 &  7:41:34.5 & $-30$:59:16 & 15.10 &  1.11 & 12.59 &  0.57 &  0.04 &  2.51 &   17.3 &  \\
 2541 &  7:41:18.5 & $-30$:51:46 & 15.11 &  1.37 & 11.83 &  0.79 &  0.15 &  3.28 &   76.6 &  \\
  329 &  7:41:47.5 & $-30$:46:50 & 15.14 &  1.61 & 11.33 &  0.92 &  0.22 &  3.81 &  121.5 &  \\
 1962 &  7:41:25.3 & $-31$:04:38 & 15.14 &  1.31 & 12.09 &  0.69 &  0.14 &  3.05 &  102.6 &  \\
 6474 &  7:40:30.2 & $-30$:48:49 & 15.14 &  1.00 & 12.94 &  0.42 &  0.04 &  2.20 &   38.1 &  \\
 2954 &  7:41:13.9 & $-30$:43:58 & 15.14 &  1.54 & 11.45 &  0.85 &  0.17 &  3.69 &  108.4 &  \\
 9186 &  7:39:56.8 & $-30$:55:47 & 15.16 &  1.04 & 13.03 &  0.44 &  0.05 &  2.13 &   46.7 &  \\
 2490 &  7:41:18.9 & $-30$:58:35 & 15.19 &  1.64 & 11.67 &  0.80 &  0.18 &  3.52 &   19.7 &  \\
10138 &  7:39:44.9 & $-31$:00:38 & 15.19 &  1.63 & 11.19 &  0.98 &  0.21 &  4.00 &  127.2 &  \\
  529 &  7:41:45.0 & $-30$:40:51 & 15.21 &  1.41 & 11.87 &  0.79 &  0.22 &  3.34 &  101.5 &  \\
10906 &  7:39:35.8 & $-31$:03:33 & 15.21 &  1.21 & 12.32 &  0.64 &  0.17 &  2.89 &   57.4 &  \\
 5658 &  7:40:41.0 & $-30$:44:20 & 15.23 &  1.26 & 12.19 &  0.73 &  0.14 &  3.04 &   99.8 &  \\
 7342 &  7:40:19.8 & $-30$:46:03 & 15.23 &  1.27 & 12.36 &  0.69 &  0.13 &  2.87 &   61.1 &  \\
 7420 &  7:40:18.9 & $-30$:48:40 & 15.25 &  1.10 & 12.52 &  0.67 &  0.19 &  2.73 &   98.8 &  \\
 9820 &  7:39:49.6 & $-30$:49:56 & 15.25 &  1.17 & 12.55 &  0.60 &  0.10 &  2.70 &   72.0 &  \\
 3713 &  7:41:03.6 & $-31$:00:35 & 15.27 &  1.16 & 12.60 &  0.63 &  0.15 &  2.67 &   42.4 &  \\
11233 &  7:39:33.1 & $-30$:49:04 & 15.29 &  1.80 & 13.55 &  0.52 &  0.07 &  1.74 &   32.4 &  2 \\
 6189 &  7:40:33.2 & $-31$:01:07 & 15.29 &  1.10 & 12.73 &  0.62 &  0.06 &  2.56 &   63.6 &  \\
 4166 &  7:40:58.3 & $-30$:54:09 & 15.30 &  1.46 & 11.79 &  0.89 &  0.14 &  3.51 &  153.4 &  \\
 2223 &  7:41:22.1 & $-30$:59:31 & 15.30 &  1.03 & 13.11 &  0.40 &  0.16 &  2.19 &   51.3 &  \\
 5296 &  7:40:45.6 & $-30$:41:13 & 15.32 &  1.27 & 12.27 &  0.79 &  0.13 &  3.05 &   66.2 &  \\
 2213 &  7:41:22.0 & $-31$:04:09 & 15.33 &  1.52 & 11.78 &  0.82 &  0.14 &  3.55 &   43.7 &  \\
10815 &  7:39:37.6 & $-30$:40:07 & 15.33 &  1.64 & 11.35 &  1.04 &  0.20 &  3.98 &   99.8 &  \\
 4326 &  7:40:57.1 & $-30$:41:35 & 15.35 &  1.33 & 12.15 &  0.74 &  0.16 &  3.20 &  100.7 &  \\
 2958 &  7:41:12.9 & $-31$:04:05 & 15.35 &  1.40 & 12.06 &  0.78 &  0.16 &  3.29 &   92.6 &  \\
 4009 &  7:41:00.7 & $-30$:49:18 & 15.35 &  1.08 & 12.90 &  0.53 &  0.10 &  2.45 &   47.4 &  \\
 6282 &  7:40:31.8 & $-31$:04:59 & 15.36 &  1.41 & 11.99 &  0.82 &  0.16 &  3.37 &   95.8 &  \\
 9954 &  7:39:48.4 & $-30$:41:49 & 15.37 &  1.04 & 13.08 &  0.46 &  0.08 &  2.29 &   39.9 &  \\
 1214 &  7:41:35.2 & $-31$:00:07 & 15.37 &  1.43 & 12.13 &  0.76 &  0.15 &  3.24 &  128.4 &  \\
 9392 &  7:39:55.0 & $-30$:43:21 & 15.37 &  1.33 & 12.19 &  0.79 &  0.12 &  3.18 &  106.5 &  \\
 1759 &  7:41:28.9 & $-30$:47:13 & 15.37 &  1.53 & 11.73 &  0.89 &  0.27 &  3.64 &  137.3 &  \\
  462 &  7:41:45.8 & $-30$:43:42 & 15.38 &  1.56 & 11.73 &  0.86 &  0.21 &  3.65 &  129.2 &  \\
 2679 &  7:41:16.4 & $-31$:03:33 & 15.39 &  1.38 & 12.18 &  0.70 &  0.14 &  3.21 &  121.3 &  \\
 2809 &  7:41:15.2 & $-30$:56:14 & 15.40 &  1.39 & 12.18 &  0.74 &  0.15 &  3.22 &   93.5 &  \\
 7010 &  7:40:23.3 & $-31$:01:02 & 15.42 &  1.34 & 12.18 &  0.76 &  0.14 &  3.24 &  101.8 &  \\
 8989 &  7:39:58.9 & $-31$:03:41 & 15.42 &  1.04 & 12.81 &  0.60 &  0.45 &  2.61 &   24.6 &  3 \\
  916 &  7:41:40.0 & $-30$:40:31 & 15.43 &  1.25 & 12.59 &  0.67 &  0.12 &  2.84 &   25.4 &  \\
 6737 &  7:40:27.5 & $-30$:44:45 & 15.44 &  1.21 & 12.84 &  0.61 &  0.06 &  2.60 &   55.3 &  \\
 5220 &  7:40:46.1 & $-30$:48:12 & 15.44 &  1.46 & 11.97 &  0.89 &  0.21 &  3.47 &   37.5 &  \\
 7249 &  7:40:19.6 & $-31$:08:47 & 15.44 &  1.21 & 12.66 &  0.62 &  0.07 &  2.78 &   57.2 &  \\
 2507 &  7:41:19.3 & $-30$:44:16 & 15.45 &  1.56 & 11.75 &  0.89 &  0.22 &  3.70 &  107.3 &  \\
 9987 &  7:39:46.7 & $-31$:05:27 & 15.46 &  1.34 & 12.26 &  0.77 &  0.17 &  3.20 &   85.6 &  \\
10026 &  7:39:45.9 & $-31$:08:48 & 15.46 &  1.37 & 12.23 &  0.74 &  0.17 &  3.23 &  110.9 &  \\
10205 &  7:39:44.2 & $-31$:01:27 & 15.47 &  1.36 & 12.22 &  0.75 &  0.17 &  3.25 &  107.1 &  \\
 9004 &  7:39:58.8 & $-31$:01:02 & 15.47 &  1.22 & 12.57 &  0.72 &  0.12 &  2.90 &   71.4 &  \\
 6725 &  7:40:27.8 & $-30$:42:10 & 15.48 &  1.80 & 11.16 &  1.06 &  0.21 &  4.32 &  112.2 &  \\
 4058 &  7:41:00.2 & $-30$:44:35 & 15.49 &  1.35 & 12.27 &  0.77 &  0.11 &  3.22 &   83.8 &  \\
  100 &  7:41:50.8 & $-30$:41:21 & 15.49 &  1.49 & 12.03 &  0.79 &  0.18 &  3.46 &   63.2 &  \\
 6643 &  7:40:28.8 & $-30$:43:43 & 15.50 &  1.90 & 12.39 &  0.81 &  0.16 &  3.11 &   67.2 &  4 \\
 8257 &  7:40:08.8 & $-30$:47:15 & 15.52 &  1.31 & 12.37 &  0.82 &  0.15 &  3.15 &   99.3 &  5 \\
 9917 &  7:39:48.7 & $-30$:45:01 & 15.53 &  1.43 & 12.14 &  0.80 &  0.12 &  3.39 &   41.6 &  \\
10923 &  7:39:37.1 & $-30$:40:53 & 15.54 &  1.51 & 12.05 &  0.82 &  0.16 &  3.49 &   72.8 &  \\
10877 &  7:39:37.6 & $-30$:40:07 & 15.58 &  1.77 & 11.35 &  1.04 &  0.20 &  4.23 &  120.2 &  \\
 3897 &  7:41:01.3 & $-31$:02:03 & 15.59 &  1.14 & 12.98 &  0.54 &  0.12 &  2.61 &   37.1 &  \\
 3242 &  7:41:09.7 & $-30$:50:45 & 15.60 &  1.05 & 13.13 &  0.59 &  0.10 &  2.47 &   33.5 &  \\
 6019 &  7:40:35.5 & $-30$:59:16 & 15.60 &  1.47 & 12.15 &  0.77 &  0.15 &  3.45 &   65.0 &  \\
 4523 &  7:40:54.5 & $-30$:42:19 & 15.61 &  1.28 & 12.59 &  0.70 &  0.12 &  3.02 &  116.5 &  \\
 4167 &  7:40:58.6 & $-30$:46:51 & 15.64 &  1.28 & 12.55 &  0.73 &  0.13 &  3.09 &   87.6 &  \\
 2617 &  7:41:17.6 & $-30$:52:00 & 15.70 &  1.70 & 11.71 &  0.98 &  0.21 &  3.99 &  109.7 &  \\
 2121 &  7:41:23.5 & $-30$:58:14 & 15.72 &  1.46 & 12.34 &  0.75 &  0.12 &  3.38 &  105.4 &  \\
  767 &  7:41:41.5 & $-30$:48:06 & 15.73 &  1.02 & 13.45 &  0.48 &  0.12 &  2.28 &   56.4 &  \\
 1295 &  7:41:35.0 & $-30$:41:18 & 15.73 &  1.03 & 13.43 &  0.45 &  0.10 &  2.30 &   33.9 &  \\
11084 &  7:39:34.7 & $-30$:46:39 & 15.74 &  1.39 & 12.38 &  0.79 &  0.18 &  3.36 &  101.3 &  \\
 5386 &  7:40:44.4 & $-30$:42:24 & 15.74 &  1.54 & 12.06 &  0.85 &  0.13 &  3.68 &  137.0 &  \\
 5967 &  7:40:35.4 & $-31$:00:26 & 15.74 &  1.18 & 11.74 &  0.36 &  0.07 &  4.00 &   49.7 &  6 \\
 5631 &  7:40:41.2 & $-30$:46:25 & 15.74 &  1.11 & 13.44 &  0.60 &  0.05 &  2.30 &   76.8 &  \\
 2055 &  7:41:24.1 & $-31$:05:25 & 15.75 &  1.96 & 11.07 &  1.14 &  0.26 &  4.68 &  118.8 &  \\
 8280 &  7:40:07.7 & $-30$:59:49 & 15.76 &  1.28 & 12.68 &  0.71 &  0.10 &  3.08 &  122.2 &  \\
 2199 &  7:41:23.3 & $-30$:41:00 & 15.77 &  1.55 & 12.09 &  0.86 &  0.21 &  3.68 &   78.0 &  \\
 3956 &  7:41:01.4 & $-30$:43:24 & 15.78 &  1.47 & 12.22 &  0.89 &  0.15 &  3.56 &   93.1 &  \\
 8070 &  7:40:11.4 & $-30$:39:32 & 15.79 &  1.07 & 13.55 &  0.44 &  0.07 &  2.24 &   51.1 &  \\
 8978 &  7:39:59.4 & $-30$:55:13 & 15.81 &  1.28 & 12.80 &  0.69 &  0.09 &  3.01 &   65.4 &  \\
  724 &  7:41:42.2 & $-30$:43:37 & 15.81 &  1.62 & 11.98 &  0.95 &  0.17 &  3.84 &   75.3 &  \\
10787 &  7:39:38.4 & $-30$:45:46 & 15.82 &  1.44 & 12.41 &  0.77 &  0.10 &  3.41 &  102.7 &  \\
 9577 &  7:39:52.4 & $-30$:48:53 & 15.84 &  1.41 & 12.50 &  0.80 &  0.18 &  3.34 &   59.4 &  \\
 8631 &  7:40:04.1 & $-30$:44:14 & 15.86 &  1.37 & 12.64 &  0.73 &  0.08 &  3.22 &  116.7 &  \\
 6950 &  7:40:23.8 & $-31$:01:32 & 15.86 &  1.40 & 12.49 &  0.85 &  0.17 &  3.37 &  124.5 &  \\
 7235 &  7:40:19.9 & $-31$:04:31 & 15.95 &  1.38 & 12.66 &  0.78 &  0.12 &  3.29 &   64.9 &  \\
 8874 &  7:39:60.0 & $-31$:04:40 & 15.95 &  1.36 & 12.66 &  0.79 &  0.18 &  3.29 &   85.2 &  \\
 5472 &  7:40:43.2 & $-30$:45:40 & 15.98 &  1.42 & 12.57 &  0.82 &  0.14 &  3.41 &  102.5 &  \\
 2927 &  7:41:13.9 & $-30$:51:29 & 16.00 &  1.28 & 13.05 &  0.66 &  0.13 &  2.95 &   50.7 &  \\
 6587 &  7:40:28.1 & $-31$:05:11 & 16.02 &  1.57 & 12.32 &  0.86 &  0.18 &  3.70 &  142.5 &  \\
  294 &  7:41:48.0 & $-30$:43:40 & 16.04 &  1.21 & 13.20 &  0.72 &  0.17 &  2.84 &   24.1 &  \\
 1084 &  7:41:37.1 & $-30$:56:36 & 16.04 &  1.29 & 13.29 &  0.77 &  0.14 &  2.75 &   91.6 &  \\
10055 &  7:39:46.6 & $-30$:51:42 & 16.10 &  1.24 & 13.31 &  0.65 &  0.11 &  2.79 &   49.5 &  \\
11459 &  7:39:29.9 & $-30$:56:18 & 16.11 &  1.35 & 13.25 &  0.74 &  0.18 &  2.86 &   55.3 &  \\
 5196 &  7:40:46.7 & $-30$:39:35 & 16.11 &  1.34 & 12.96 &  0.72 &  0.11 &  3.15 &  102.8 &  \\
 9405 &  7:39:54.8 & $-30$:45:34 & 16.12 &  1.62 & 12.33 &  0.90 &  0.09 &  3.79 &   97.8 &  \\
  396 &  7:41:45.9 & $-31$:05:37 & 16.13 &  1.25 & 13.34 &  0.63 &  0.13 &  2.79 &   45.2 &  \\
 9912 &  7:39:47.8 & $-30$:59:31 & 16.13 &  1.07 & 13.69 &  0.55 &  0.14 &  2.44 &   33.5 &  \\
 9978 &  7:39:47.2 & $-30$:56:56 & 16.13 &  1.25 & 13.16 &  0.76 &  0.26 &  2.97 &   16.6 &  \\
11436 &  7:39:30.9 & $-30$:45:06 & 16.13 &  1.10 & 13.83 &  0.61 &  0.23 &  2.30 &   74.7 &  \\
 6486 &  7:40:30.3 & $-30$:43:53 & 16.15 &  1.11 & 13.87 &  0.52 &  0.13 &  2.28 &   57.5 &  \\
 1179 &  7:41:35.3 & $-31$:05:56 & 16.15 &  1.13 & 14.48 &  0.44 &  0.03 &  1.67 &   72.1 &  7 \\
 8569 &  7:40:04.8 & $-30$:43:27 & 16.16 &  1.33 & 13.09 &  0.68 &  0.12 &  3.07 &   93.6 &  \\
11107 &  7:39:33.5 & $-31$:03:51 & 16.19 &  1.44 & 12.71 &  0.85 &  0.20 &  3.48 &  125.8 &  \\
 1258 &  7:41:35.4 & $-30$:39:22 & 16.19 &  1.36 & 12.95 &  0.73 &  0.13 &  3.24 &   99.6 &  \\
 7413 &  7:40:18.1 & $-31$:02:55 & 16.22 &  1.29 & 13.23 &  0.73 &  0.13 &  2.99 &   56.5 &  \\
 1148 &  7:41:36.0 & $-31$:01:16 & 16.23 &  1.01 & 13.72 &  0.87 &  0.13 &  2.51 &   56.8 &  8 \\
 7387 &  7:40:19.6 & $-30$:42:09 & 16.24 &  1.43 & 13.28 &  0.78 &  0.16 &  2.96 &   29.9 &  \\
 6645 &  7:40:28.9 & $-30$:40:30 & 16.28 &  1.02 & 13.99 &  0.53 &  0.03 &  2.29 &   74.0 &  \\
 8136 &  7:40:09.2 & $-31$:04:04 & 16.29 &  1.15 & 13.47 &  0.69 &  0.17 &  2.82 &   84.6 &  \\
 3744 &  7:41:03.6 & $-30$:51:27 & 16.29 &  1.05 & 13.90 &  0.48 &  0.15 &  2.39 &  103.8 &  \\
 4730 &  7:40:52.0 & $-30$:48:16 & 16.29 &  1.32 & 13.11 &  0.75 &  0.15 &  3.18 &   82.3 &  \\
 5665 &  7:40:39.8 & $-31$:04:43 & 16.30 &  1.35 & 13.10 &  0.83 &  0.22 &  3.20 &  125.3 &  9 \\
 9124 &  7:39:57.2 & $-31$:01:24 & 16.31 &  1.39 & 12.92 &  0.84 &  0.19 &  3.39 &  126.4 &  \\
 6704 &  7:40:27.0 & $-31$:00:18 & 16.32 &  1.42 & 12.87 &  0.83 &  0.18 &  3.45 &   91.9 &  \\
 2285 &  7:41:21.2 & $-30$:59:48 & 16.32 &  1.48 & 12.92 &  0.76 &  0.17 &  3.40 &  103.1 &  \\
 1436 &  7:41:32.7 & $-30$:53:52 & 16.33 &  1.03 & 14.11 &  0.40 &  0.00 &  2.22 &   70.2 &  \\
 1636 &  7:41:30.2 & $-30$:57:20 & 16.36 &  1.38 & 13.09 &  0.78 &  0.17 &  3.27 &  110.6 &  \\
 2727 &  7:41:16.1 & $-30$:56:36 & 16.36 &  1.04 & 14.16 &  0.49 &  0.05 &  2.20 &  105.1 &  \\
 6232 &  7:40:32.4 & $-31$:04:12 & 16.37 &  1.38 & 13.03 &  0.76 &  0.06 &  3.34 &  112.3 &  \\
 8641 &  7:40:03.2 & $-30$:56:57 & 16.37 &  1.27 & 13.37 &  0.67 &  0.10 &  3.00 &   79.1 &  \\
 8887 &  7:39:59.7 & $-31$:06:14 & 16.38 &  1.35 & 13.23 &  0.74 &  0.18 &  3.15 &   39.8 &  \\
  558 &  7:41:43.8 & $-31$:02:12 & 16.39 &  1.55 & 12.76 &  0.78 &  0.14 &  3.63 &   58.3 &  \\
 2864 &  7:41:14.5 & $-30$:53:06 & 16.40 &  1.29 & 13.32 &  0.71 &  0.15 &  3.08 &  116.3 &  \\
11563 &  7:39:28.9 & $-30$:51:45 & 16.40 &  1.33 & 13.36 &  0.68 &  0.11 &  3.04 &   76.6 &  \\
11399 &  7:39:31.1 & $-30$:49:41 & 16.41 &  1.09 & 14.02 &  0.66 &  0.12 &  2.39 &   39.2 &  \\
 2969 &  7:41:12.8 & $-31$:02:22 & 16.42 &  1.45 & 13.07 &  0.74 &  0.16 &  3.35 &  127.9 &  \\
 8346 &  7:40:07.6 & $-30$:45:02 & 16.44 &  1.62 & 13.10 &  0.74 &  0.15 &  3.34 &   42.0 &  \\
 7600 &  7:40:16.9 & $-30$:42:34 & 16.45 &  1.07 & 14.02 &  0.55 &  0.15 &  2.43 &   19.5 &  \\
 3004 &  7:41:12.7 & $-30$:54:42 & 16.45 &  1.52 & 13.24 &  0.71 &  0.04 &  3.21 &   43.8 &  \\
 2717 &  7:41:16.6 & $-30$:47:15 & 16.46 &  1.45 & 13.15 &  0.89 &  0.21 &  3.31 &   66.9 &  \\
 9200 &  7:39:56.2 & $-31$:02:27 & 16.47 &  1.33 & 13.21 &  0.77 &  0.15 &  3.26 &   79.5 &  \\
  955 &  7:41:39.1 & $-30$:48:16 & 16.47 &  1.00 & 14.13 &  0.56 &  0.25 &  2.34 &   44.8 &  \\
 3751 &  7:41:03.6 & $-31$:00:35 & 16.47 &  1.48 & 12.60 &  0.63 &  0.15 &  3.87 &  102.2 &  \\
 5547 &  7:40:42.5 & $-30$:41:25 & 16.50 &  1.03 & 14.09 &  0.72 &  0.21 &  2.41 &   38.1 &  \\
\enddata
\tablenotetext{a}{From 2MASS}
\tablecomments{1. Uncertain photometry \& identification. 2. Uncertain photometry.
3. Uncertain 2MASS photometry. 4. $K$ = 7.3 mag star 8\arcsec\ away. 5. $K$ magnitude
is uncertain. 6. $K$ = 6.5 mag star 13\arcsec\ away. 7. Uncertain photometry \&
identification. 8. Uncertain photometry \& identification: $K$ = 14.4 mag star
4\arcsec\ away. 9. Uncertain photometry: $K$ = 14.2 mag 4\arcsec\ away.}
\end{deluxetable}

\clearpage 

\begin{deluxetable}{lrrlllll}
\tabletypesize{\footnotesize}
\tablenum{4}
\tablecolumns{8}
\tablecaption{Field Giants in the Southern Warp: 
Spectroscopic data \label{tab:observations}}
\tablewidth{0pt}
\tablehead{
  \colhead{Star} & 
  \colhead{$\ell$} & 
  \colhead{$b$} &
  \colhead{HJD} &   
  \colhead{Exp. Time} & 
  \colhead{S/N} & 
  \colhead{$V_{\rm rad}$} &
  \colhead{$V_{\rm GSR}$} \\
  & & & 
  \colhead{$-2400000$} & 
  \colhead{(min)} &   &
  \colhead{(\kms)} &
  \colhead{(\kms)} 
}
\startdata
9060 &  245.71 & $-4.25$ & 51203.84616 & 2$\times$60 & 59 & $+132.9 \pm 1.5$ & $-86.0$ \\
 & & & 51204.85843 & 2$\times$60 & 49 & & \\ 
 & & & 51205.85450 & 2$\times$60 & 47 & & \\ 
7512 & 245.55 & $-4.08$ & 50834.65961 & 5$\times$60 & 77 & $+139.8 \pm 0.5$ & $-78.8$ \\
4166 & 245.77 & $-4.04$ & 50833.68964 & 5$\times$60 & 71 & $+156.2 \pm 0.5$ & $-62.7$ \\
\enddata
\end{deluxetable}

\clearpage

\begin{deluxetable}{lrrrrrrr}
\tablenum{5}
\tablecolumns{8}
\tablewidth{0pt}
\tabletypesize{\footnotesize}
\tablecaption{Atmospheric Parameters \label{tab:parameters}}
\tablehead{
  \colhead{Star} & 
  \colhead{[Fe/H]\tablenotemark{a}} & 
  \colhead{$T_{\rm eff}$\tablenotemark{b}} & 
  \colhead{log g\tablenotemark{b}} & 
  \colhead{$T_{\rm eff}$\tablenotemark{c}} &   
  \colhead{log g\tablenotemark{c}} & 
  \colhead{$V_{\rm turb}$} &
  \colhead{[Fe/H]\tablenotemark{c}} 
}
\startdata
 \noalign{\vskip +0.5ex}
 9060 & $-0.6$  & 5480 & 1.4 & 5250 & 3.0 & 1.5 & $-0.48$ \\
 7512 & $-0.6$  & 4750 & 1.0 & 4590 & 2.25 & 1.4 & $-0.40$ \\
 4166 & $-0.6$  & 4860 & 1.0 & 4590 & 2.2 & 1.6 & $-0.42$ \\
\enddata
\tablenotetext{a}{Adopted value based on the Galactic metallicity gradient.}
\tablenotetext{b}{Estimates obtained using the reddening estimate
from Schlegel et al.\ (1998) and distance estimates based on
the radial velocity and an adopted circular Galactic rotation.}
\tablenotetext{c}{Quantities derived using the spectroscopic methods
described in the text.}

\end{deluxetable}

\clearpage

\begin{deluxetable}{lccrcclccrcclccrc} 
\tablenum{6}
\tabletypesize{\tiny}
\rotate
\tablecolumns{17} 
\tablewidth{0pc} 
\tablecaption{Stellar Atomic Line Data \label{tab:lines}}
\tablehead{ 
\colhead{$\lambda$(\AA)} &
\colhead{Species} &
\colhead{LEP(eV)} &
\colhead{log $gf$} &
\colhead{Source} &
\colhead{} &
\colhead{$\lambda$(\AA)} &
\colhead{Species} &
\colhead{LEP(eV)} &
\colhead{log $gf$} &
\colhead{Source} &
\colhead{} &
\colhead{$\lambda$(\AA)} &
\colhead{Species} &
\colhead{LEP(eV)} &
\colhead{log $gf$} &
\colhead{Source}
}
\startdata
6300.30 & 8.0 & 0.00 & $-$9.717 & AG04 & & 5855.09 & 26.0 & 4.60 & $-$1.547 & PS03 & & 6575.02 & 26.0 & 2.59 & $-$2.727 & OXF \\
6363.78 & 8.0 & 0.02 & $-$10.185 & AG04 & & 5856.10 & 26.0 & 4.29 & $-$1.640 & PS03 & & 6581.21 & 26.0 & 1.50 & $-$4.680 & AN00 \\
5688.19 & 11.0 & 2.11 & $-$0.420 & RC02 & & 5858.79 & 26.0 & 4.22 & $-$2.260 & PS03 & & 6609.11 & 26.0 & 2.56 & $-$2.692 & OXF \\
6154.23 & 11.0 & 2.10 & $-$1.530 & RC02 & & 5909.97 & 26.0 & 3.21 & $-$2.640 & OXF & & 6648.08 & 26.0 & 1.01 & $-$5.918 & OXF \\
6160.75 & 11.0 & 2.10 & $-$1.230 & RC02 & & 5916.25 & 26.0 & 2.45 & $-$2.994 & OXF & & 6699.16 & 26.0 & 4.59 & $-$2.170 & OXF \\
6318.72 & 12.0 & 5.11 & $-$1.970 & RC02 & & 5927.80 & 26.0 & 4.65 & $-$1.090 & PS03 & & 6739.52 & 26.0 & 1.56 & $-$4.820 & OXF \\
6319.24 & 12.0 & 5.11 & $-$2.220 & RC02 & & 5933.80 & 26.0 & 4.64 & $-$2.230 & PS03 & & 6750.15 & 26.0 & 2.42 & $-$2.621 & OXF \\
6965.41 & 12.0 & 5.75 & $-$1.510 & KB95 & & 5956.69 & 26.0 & 0.86 & $-$4.608 & OXF & & 6786.86 & 26.0 & 4.19 & $-$1.850 & TF00 \\
7387.69 & 12.0 & 5.75 & $-$0.870 & RC02 & & 6012.21 & 26.0 & 2.22 & $-$4.070 & OXF & & 6810.26 & 26.0 & 4.60 & $-$1.000 & OXF \\
6696.02 & 13.0 & 3.14 & $-$1.340 & RC02 & & 6019.36 & 26.0 & 3.57 & $-$3.360 & PS03 & & 6971.93 & 26.0 & 3.02 & $-$3.390 & OXF \\
6698.67 & 13.0 & 3.14 & $-$1.640 & RC02 & & 6027.05 & 26.0 & 4.07 & $-$1.106 & OXF & & 7112.17 & 26.0 & 2.99 & $-$3.044 & OXF \\
7835.31 & 13.0 & 4.02 & $-$0.470 & LUCK & & 6054.08 & 26.0 & 4.37 & $-$2.310 & PS03 & & 7223.66 & 26.0 & 3.01 & $-$2.269 & OXF \\
7836.13 & 13.0 & 4.02 & $-$0.310 & RC02 & & 6105.13 & 26.0 & 4.55 & $-$2.050 & PS03 & & 5991.38 & 26.1 & 3.15 & $-$3.557 & BB91 \\
5690.43 & 14.0 & 4.93 & $-$1.751 & SOLAR & & 6120.24 & 26.0 & 0.91 & $-$5.970 & OXF & & 6084.11 & 26.1 & 3.20 & $-$3.808 & BB91 \\
5793.07 & 14.0 & 4.93 & $-$1.843 & SOLAR & & 6145.42 & 26.0 & 3.37 & $-$3.600 & TF00 & & 6149.26 & 26.1 & 3.89 & $-$2.724 & BB91 \\
6125.02 & 14.0 & 5.61 & $-$1.506 & SOLAR & & 6151.62 & 26.0 & 2.17 & $-$3.299 & OXF & & 6247.56 & 26.1 & 3.89 & $-$2.329 & BB91 \\
6145.01 & 14.0 & 5.62 & $-$1.362 & SOLAR & & 6157.73 & 26.0 & 4.08 & $-$1.320 & TF00 & & 6369.46 & 26.1 & 2.89 & $-$4.250 & BB91 \\
6155.13 & 14.0 & 5.62 & $-$0.786 & SOLAR & & 6159.38 & 26.0 & 4.61 & $-$1.970 & PS03 & & 6416.92 & 26.1 & 3.89 & $-$2.740 & BB91 \\
6166.44 & 20.0 & 2.52 & $-$1.142 & OXF & & 6165.36 & 26.0 & 4.14 & $-$1.490 & OXF & & 6432.68 & 26.1 & 2.89 & $-$3.708 & BB91 \\
6169.04 & 20.0 & 2.52 & $-$0.797 & OXF & & 6173.34 & 26.0 & 2.22 & $-$2.880 & OXF & & 6456.38 & 26.1 & 3.90 & $-$2.075 & BB91 \\
6169.56 & 20.0 & 2.53 & $-$0.478 & OXF & & 6180.20 & 26.0 & 2.73 & $-$2.637 & OXF & & 6189.00 & 27.0 & 1.71 & $-$2.450 & PN00 \\
6455.60 & 20.0 & 2.52 & $-$1.290 & OXF & & 6200.31 & 26.0 & 2.61 & $-$2.437 & OXF & & 6455.03 & 27.0 & 3.63 & $-$0.250 & PN00 \\
6064.63 & 22.0 & 1.05 & $-$1.888 & OXF & & 6219.28 & 26.0 & 2.20 & $-$2.433 & OXF & & 6632.45 & 27.0 & 2.28 & $-$2.000 & PN00 \\
6091.17 & 22.0 & 2.27 & $-$0.367 & OXF & & 6229.23 & 26.0 & 2.84 & $-$2.846 & OXF & & 5846.99 & 28.0 & 1.68 & $-$3.210 & RC02 \\
6312.24 & 22.0 & 1.46 & $-$1.496 & OXF & & 6232.64 & 26.0 & 3.65 & $-$1.283 & OXF & & 6086.28 & 28.0 & 4.26 & $-$0.515 & RC02 \\
6336.10 & 22.0 & 1.44 & $-$1.687 & OXF & & 6246.32 & 26.0 & 3.60 & $-$0.894 & OXF & & 6175.37 & 28.0 & 4.09 & $-$0.535 & RC02 \\
6013.53 & 25.0 & 3.07 & $-$0.250 & PM00 & & 6265.13 & 26.0 & 2.17 & $-$2.550 & OXF & & 6177.24 & 28.0 & 1.83 & $-$3.510 & RC02 \\
6016.67 & 25.0 & 3.08 & $-$0.220 & PM00 & & 6270.22 & 26.0 & 2.86 & $-$2.500 & OXF & & 6204.60 & 28.0 & 4.09 & $-$1.140 & RC02 \\
6021.80 & 25.0 & 3.08 & 0.030 & PM00 & & 6271.28 & 26.0 & 3.33 & $-$2.703 & AN00 & & 6635.12 & 28.0 & 4.42 & $-$0.828 & RC02 \\
5618.63 & 26.0 & 4.21 & $-$1.292 & OXF & & 6297.79 & 26.0 & 2.22 & $-$2.740 & OXF & & 6772.32 & 28.0 & 3.66 & $-$0.987 & RC02 \\
5701.55 & 26.0 & 2.56 & $-$2.216 & OXF & & 6301.50 & 26.0 & 3.65 & $-$0.766 & OXF & & 7800.00 & 37.0 & 0.00 & 0.130 & TL99 \\
5705.47 & 26.0 & 4.30 & $-$1.420 & OXF & & 6322.69 & 26.0 & 2.59 & $-$2.426 & OXF & & 6127.44 & 40.0 & 0.15 & $-$1.060 & RC02 \\
5741.85 & 26.0 & 4.25 & $-$1.689 & OXF & & 6336.82 & 26.0 & 3.68 & $-$0.916 & OXF & & 6134.55 & 40.0 & 0.00 & $-$1.280 & RC02 \\
5775.08 & 26.0 & 4.22 & $-$1.310 & OXF & & 6353.84 & 26.0 & 0.91 & $-$6.477 & OXF & & 6143.20 & 40.0 & 0.07 & $-$1.100 & RC02 \\
5778.45 & 26.0 & 2.59 & $-$3.480 & OXF & & 6355.03 & 26.0 & 2.84 & $-$2.403 & OXF & & 5853.64 & 56.1 & 0.60 & $-$1.010 & PN00 \\
5811.92 & 26.0 & 4.14 & $-$2.430 & PS03 & & 6411.65 & 26.0 & 3.65 & $-$0.734 & OXF & & 5805.77 & 57.1 & 0.13 & $-$1.560 & LB01 \\
5837.70 & 26.0 & 4.29 & $-$2.340 & TF00 & & 6469.19 & 26.0 & 4.84 & $-$0.770 & TF00 & & 6390.48 & 57.1 & 0.32 & $-$1.410 & LB01 \\
5853.16 & 26.0 & 1.49 & $-$5.280 & PS03 & & 6574.23 & 26.0 & 0.99 & $-$5.004 & OXF & & 6645.11 & 63.1 & 1.38 & 0.120 & LW01 \\
\enddata

\tablerefs{
AG04 = Asplund et al.\ ( 2004);
AN00 = Asplund et al.\ (2000);
BB91 = Biemont et al.\ (1991);
KB05 = Kurucz \& Bell (1995);
LB01 = Lawler, Bonvallet, \& Sneden (2000);
LW01 = Lawler et al.\ (2001);
LUCK = Luck (private communication);
OXF = Group at Oxford (Smith \& Raggett 1981, Blackwell et~al.\ 1979a,b; 
1980; 1982a,b; 1983; 1986a,b; 1995);
PM00 = Prochaska \& McWilliam (2000);
PN00 = Prochaska et al.\ (2000);
PS03 = Paulson, Sneden, \& Cochran (2003);
RC02 = Ram{\'{\i}}rez \& Cohen (2002);
SOLAR = Inverted Solar analysis;
TF00 = Thor{\' e}n \& Feltzing (2000);
TL99 = Tomkin \& Lambert (1999).
}

\end{deluxetable}

\clearpage

\begin{deluxetable}{lrrrrrrrrrrr} 
\tabletypesize{\footnotesize}
\tablecolumns{12} 
\tablewidth{0pc} 
\tablecaption{Mean stellar abundances \label{tab:abundances}}
\tablenum{7}
\tablehead{ 
\colhead{Species} &
\colhead{Abundance} &
\colhead{$\sigma$} &
\colhead{N} &
\colhead{} &
\colhead{Abundance} &
\colhead{$\sigma$} &
\colhead{N} &
\colhead{} &
\colhead{Abundance} &
\colhead{$\sigma$} &
\colhead{N}
}
\startdata
 \noalign{\vskip +0.5ex}
\hline
 \noalign{\vskip +0.5ex}
\colhead{} & 
\multicolumn{3}{c}{\bf 4166} & & 
\multicolumn{3}{c}{\bf 7521} & & 
\multicolumn{3}{c}{\bf 9060} \\
 \noalign{\vskip  .8ex} \hline
 \noalign{\vskip -2ex}\\
{\rm [O/Fe]}   &    0.19 & 0.14    & 2  & &    0.15 & 0.11    & 2  & &    0.44    & 0.07    & 2       \\
{\rm [Na/Fe]}  &    0.24 & 0.08    & 3  & &    0.28 & 0.18    & 3  & &    0.37    & 0.10    & 3       \\
{\rm [Mg/Fe]}  &    0.34 & 0.06    & 4  & &    0.24 & 0.04    & 4  & &    0.21    & 0.06    & 4       \\
{\rm [Al/Fe]}  &    0.30 & 0.04    & 2  & &    0.26 & 0.10    & 4  & &    0.15    & 0.03    & 4       \\
{\rm [Si/Fe]}  &    0.19 & 0.07    & 5  & &    0.18 & 0.08    & 5  & &    0.12    & 0.11    & 5       \\
{\rm [Ca/Fe]}  &    0.25 & 0.16    & 4  & &    0.29 & 0.14    & 4  & &    0.03    & 0.03    & 4       \\
{\rm [Ti/Fe]}  &    0.23 & 0.07    & 4  & &    0.33 & 0.10    & 4  & &    0.41    & 0.06    & 3       \\
{\rm [Mn/Fe]}  & $-$0.29 & 0.10    & 3  & & $-$0.23 & 0.11    & 3  & & $-$0.23    & 0.11    & 3       \\
{\rm [FeI/H]}  & $-$0.45 & 0.17    & 38 & & $-$0.43 & 0.20    & 41 & & $-$0.51    & 0.14    & 44      \\
{\rm [FeII/H]} & $-$0.49 & 0.07    & 8  & & $-$0.47 & 0.04    & 7  & & $-$0.53    & 0.14    & 8       \\
{\rm [Co/Fe]}  &    0.10 & 0.18    & 3  & &    0.16 & 0.09    & 3  & &    0.26    & 0.19    & 3       \\
{\rm [Ni/Fe]}  &    0.06 & 0.13    & 7  & &    0.07 & 0.11    & 7  & &    0.03    & 0.11    & 7       \\
{\rm [Rb/Fe]}  &    0.10 & \nodata & 1  & &    0.14 & \nodata & 1  & &    \nodata & \nodata & \nodata \\
{\rm [Zr/Fe]}  & $-$0.12 & 0.04    & 3  & & $-$0.08 & 0.17    & 3  & &    \nodata & \nodata & \nodata \\
{\rm [Ba/Fe]}  &    0.16 & \nodata & 1  & & $-$0.03 & \nodata & 1  & & $-$0.14    & \nodata & 1       \\
{\rm [La/Fe]}  &    0.36 & 0.01    & 2  & &    0.27 & 0.04    & 2  & &    0.26    & 0.07    & 2       \\
{\rm [Eu/Fe]}  &    0.46 & \nodata & 1  & &    0.39 & \nodata & 1  & &    0.66    & \nodata & 1       \\
\enddata

\end{deluxetable}

\clearpage

\begin{deluxetable}{lrrrrrrrr}
\tablenum{8}
\tablecolumns{9}
\tablewidth{0pt}
\tabletypesize{\footnotesize}
\tablecaption{Revised Distance Estimates \label{tab:distances}}
\tablehead{
  \colhead{Star} & 
  \colhead{$V-K$} & 
  \colhead{($V-K$)$_{0}$\tablenotemark{a}} & 
  \colhead{$T_{\rm eff}$\tablenotemark{a}} & 
  \colhead{E($B-V$)} & 
  \colhead{$M_{K}$\tablenotemark{b}} &
  \colhead{($m-M$)$_{0}$} &
  \colhead{d\tablenotemark{c}} &
  \colhead{$R_{\rm GC}$\tablenotemark{c}}
}
\startdata
 \noalign{\vskip +0.5ex}
 9060 & 2.94 & 1.90 & 5216 & 0.39 & $-1.6$ & 13.08 & 4.13 & 10.40 \\ 
 7512 & 3.73 & 2.50 & 4581 & 0.45 & $-3.2$ & 14.39 & 7.55 & 13.07 \\
 4166 & 3.51 & 2.50 & 4582 & 0.37 & $-3.2$ & 14.86 & 9.38 & 14.61 \\
\enddata
\tablenotetext{a}{The de-reddened $V-K$ value is that needed to produce
the $T_{\rm eff}$ value cited here, and which agrees well with the
value derived spectroscopically: see Table~\ref{tab:parameters}.}
\tablenotetext{b}{The $M_{\rm K}$ value has been derived using
the relation between this value and $T_{\rm eff}$ for the open
cluster Be~29, which has a very similar metallicity as the field stars,
as described in the text.}
\tablenotetext{c}{Distances are given in kpc.}
\end{deluxetable}
\clearpage

\begin{deluxetable}{lrrrrrr}
\tablenum{9}
\tablecolumns{7}
\tablewidth{0pt}
\tabletypesize{\footnotesize}
\tablecaption{Abundance Comparisons\tablenotemark{a} \label{tab:comparisons}}
\tablehead{
  \colhead{Element} & 
  \colhead{M67 means} & 
  \colhead{$\sigma$} &
  \colhead{Outer disk clusters\tablenotemark{b}} & 
  \colhead{$\sigma$} &
  \colhead{Warp stars} &
  \colhead{$\sigma$} 
}
\startdata
O &  $+0.08 \pm 0.01$ & 0.03 & $+0.21 \pm 0.02$ & 0.04 & $+0.26 \pm 0.09$ & 0.16 \\
Na & $+0.30 \pm 0.04$ & 0.07 & $+0.31 \pm 0.05$ & 0.11 & $+0.30 \pm 0.04$ & 0.07 \\
Mg & $+0.16 \pm 0.01$ & 0.02 & $+0.30 \pm 0.03$ & 0.07 & $+0.26 \pm 0.04$ & 0.07 \\
Al & $+0.17 \pm 0.01$ & 0.01 & $+0.21 \pm 0.02$ & 0.04 & $+0.24 \pm 0.05$ & 0.08 \\
Mg & $+0.16 \pm 0.01$ & 0.02 & $+0.30 \pm 0.03$ & 0.07 & $+0.26 \pm 0.04$ & 0.07 \\
Si & $+0.09 \pm 0.01$ & 0.02 & $+0.12 \pm 0.03$ & 0.08 & $+0.16 \pm 0.02$ & 0.04 \\
Ca & $+0.07 \pm 0.02$ & 0.03 & $+0.07 \pm 0.02$ & 0.05 & $+0.19 \pm 0.08$ & 0.14 \\
Ti & $+0.12 \pm 0.04$ & 0.06 & $+0.29 \pm 0.05$ & 0.13 & $+0.36 \pm 0.05$ & 0.09 \\
Mn & $-0.13 \pm 0.04$ & 0.06 & $-0.24 \pm 0.07$ & 0.16 & $-0.25 \pm 0.02$ & 0.03 \\
Co & $+0.02 \pm 0.01$ & 0.01 & $+0.14 \pm 0.03$ & 0.08 & $+0.17 \pm 0.05$ & 0.08 \\
Ni & $+0.08 \pm 0.02$ & 0.03 & $+0.01 \pm 0.02$ & 0.06 & $+0.05 \pm 0.01$ & 0.02 \\
Rb & $-0.27 \pm 0.03$ & 0.05 & $-0.09 \pm 0.05$ & 0.12 & $+0.12 \pm 0.02$ & 0.03 \\
         &                  &      & $-0.04 \pm 0.08$ & 0.14 &                  &      \\
         &                  &      & $-0.16 \pm 0.05$ & 0.06 &                  &      \\
Zr & $-0.28 \pm 0.02$ & 0.03 & $+0.20 \pm 0.10$ & 0.24 & $-0.10 \pm 0.02$ & 0.03 \\
         &                  &      & $+0.06 \pm 0.01$ & 0.03 &                  &      \\
         &                  &      & $+0.49 \pm 0.14$ & 0.20 &                  &      \\
Ba & $-0.02 \pm 0.02$ & 0.04 & $+0.41 \pm 0.13$ & 0.31 & $+0.00 \pm 0.09$ & 0.15 \\
         &                  &      & $+0.22 \pm 0.05$ & 0.10 &                  &      \\
         &                  &      & $+0.78 \pm 0.14$ & 0.19 &                  &      \\
La & $+0.11 \pm 0.01$ & 0.02 & $+0.42 \pm 0.11$ & 0.27 & $+0.30 \pm 0.03$ & 0.06 \\
         &                  &      & $+0.27 \pm 0.04$ & 0.07 &                  &      \\
         &                  &      & $+0.74 \pm 0.17$ & 0.24 &                  &      \\
Eu & $+0.06 \pm 0.02$ & 0.03 & $+0.29 \pm 0.06$ & 0.16 & $+0.50 \pm 0.08$ & 0.14 \\
         &                  &      & $+0.23 \pm 0.05$ & 0.10 &                  &      \\
         &                  &      & $+0.37 \pm 0.20$ & 0.28 &                  &      \\
\enddata
\tablenotetext{a}{The values in parentheses are standard errors, while the
quoted uncertainty is the error of the mean.}
\tablenotetext{b}{For the $s$-process elements Rb, Zr, Ba, and La, and
the $r$-process element Eu we provide three means for the clusters.
The first row is for all the stars in all the clusters. The second
row includes results only for Be~20 and Be~29.
The third row is for the two clusters with enhanced neutron capture
elements, Be~31 and NGC~2141.}
\end{deluxetable}

\clearpage

\clearpage

\begin{figure}
\epsscale{0.8}
\caption{A view of Warp Field 1a, with expanded views of the
regions surrounding our three primary targets. \label{fig:chart}}
\end{figure}

\clearpage

\begin{figure}
\epsscale{0.8}
\plotone{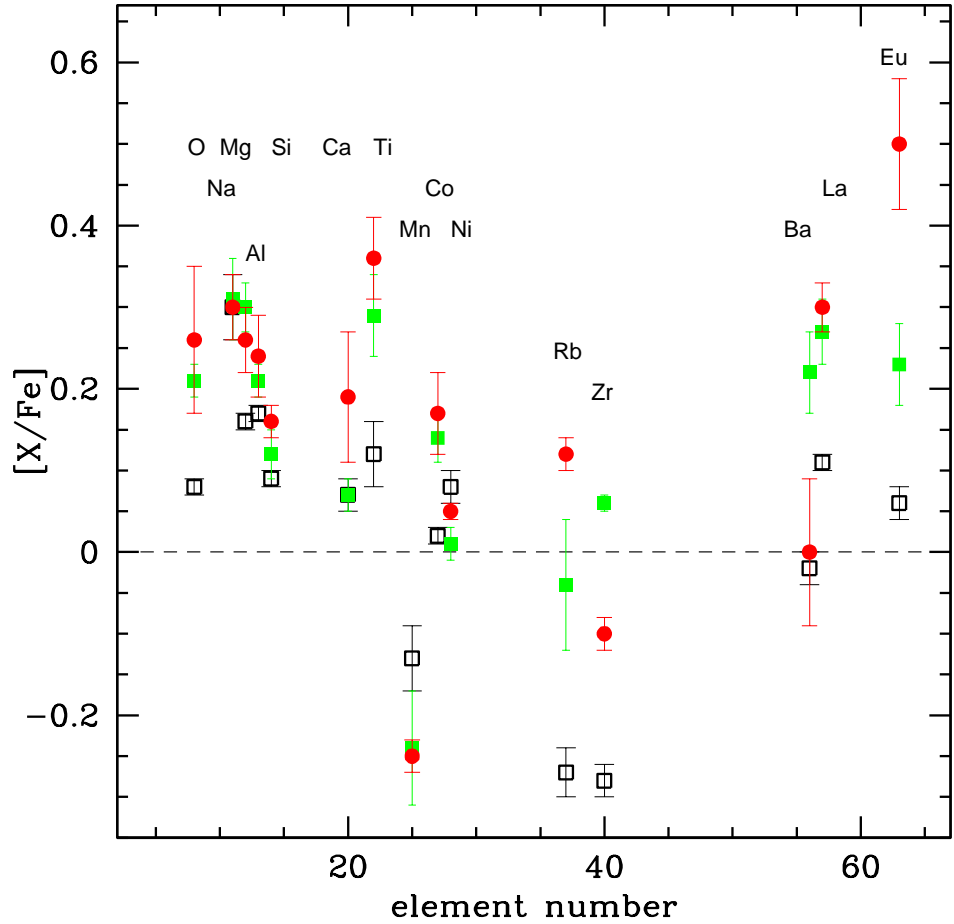}
\caption{A comparison of the mean abundances for stars from
Paper~I and this paper. Plus signs represent the three
stars in the old open cluster M67, filled green squares
are the six stars in the old open clusters Be~20, Be~29,
Be~31, and NGC~2141
in the outer Galactic disk, and the filled red circles
are the field stars analyzed in this paper. For the
elements Rb, Zr, Ba, La, and Eu we show results only
for Be~20 and Be~29, for reasons discussed in the text.
\label{fig:xfecomparisons}}
\end{figure}

\clearpage

\begin{figure}
\epsscale{0.8}
\plotone{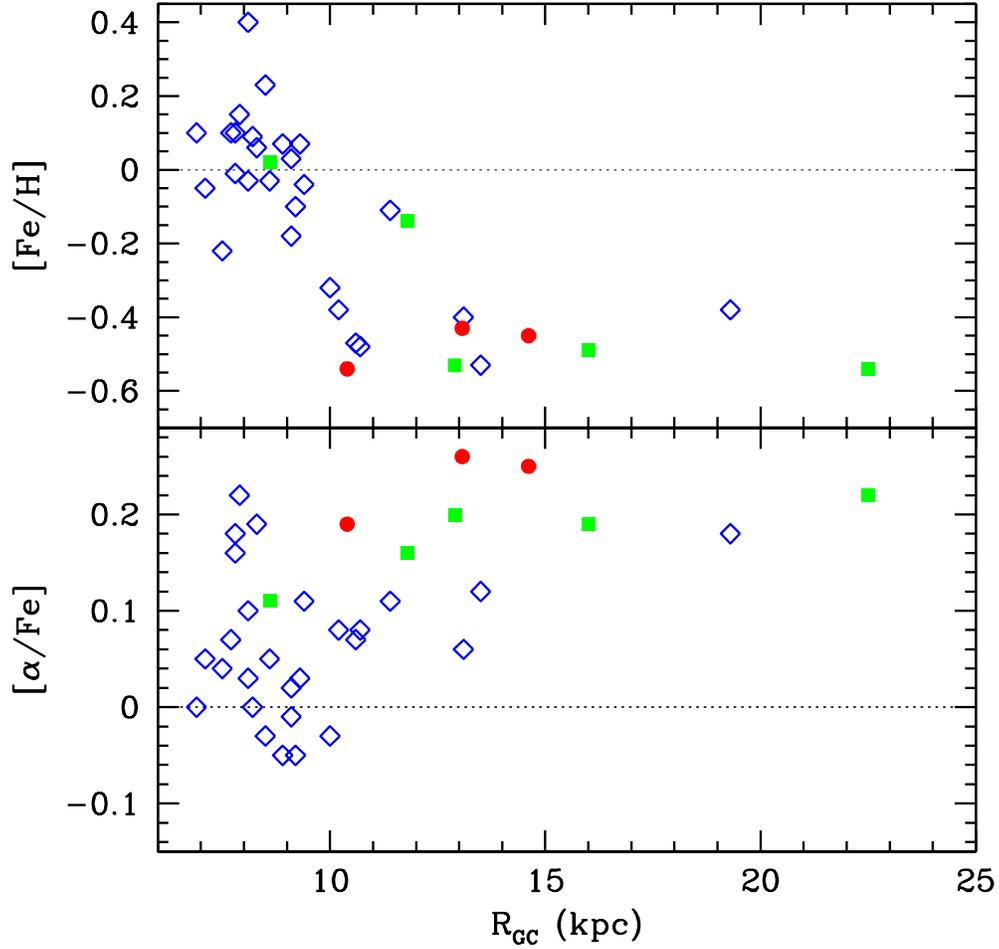}
\caption{{\em Upper:} The metallicity gradient of old open clusters from the
literature (blue diamonds), from Yong et al.\ (2005; green squares),
and the three field stars discussed in this paper (filled red circles).
{\em Lower:} The mean abundances of Mg, Si, Ca, and Ti compared to
iron, using the same symbols, as a function of Galactocentric
distance. \label{fig:rgcwarp}}
\end{figure}

\clearpage

\begin{figure}
\epsscale{0.8}
\plotone{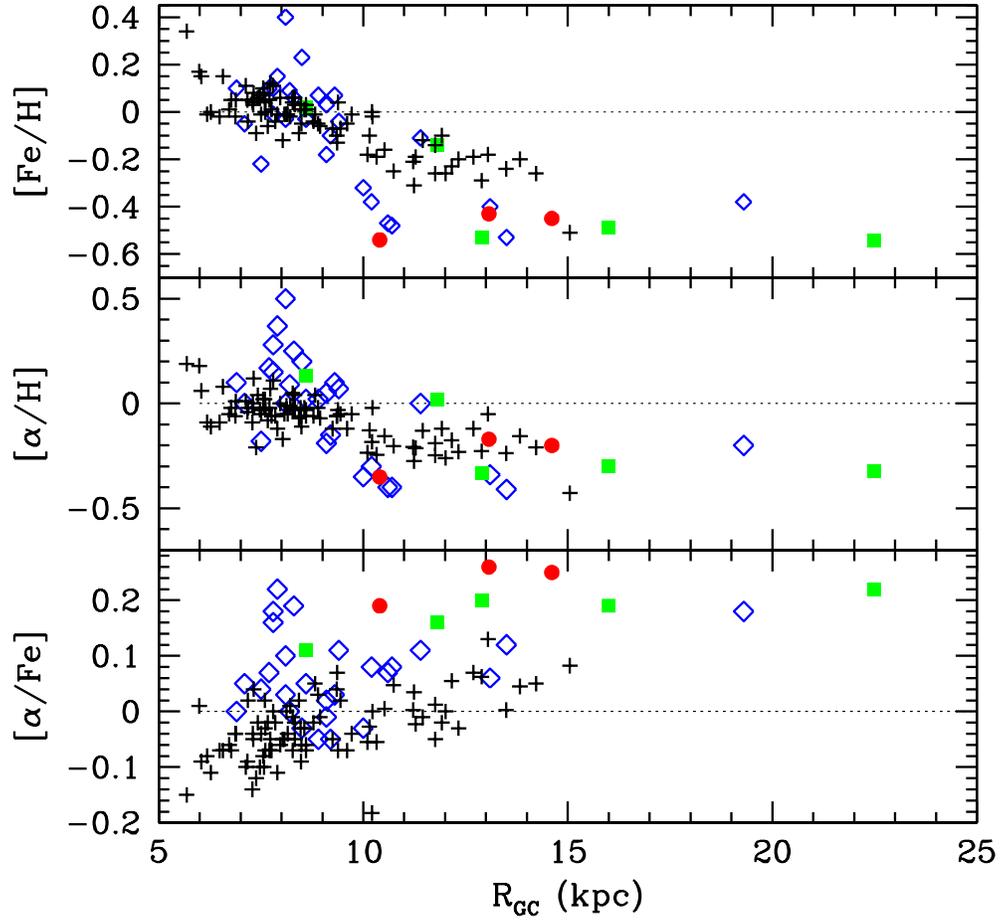}
\caption{The same as Figure~\ref{fig:rgcwarp},
but also including the results from Cepheid variables (black
plus signs). \label{fig:cepheidswarp}}
\end{figure}

\clearpage

\begin{figure}
\epsscale{0.8}
\plotone{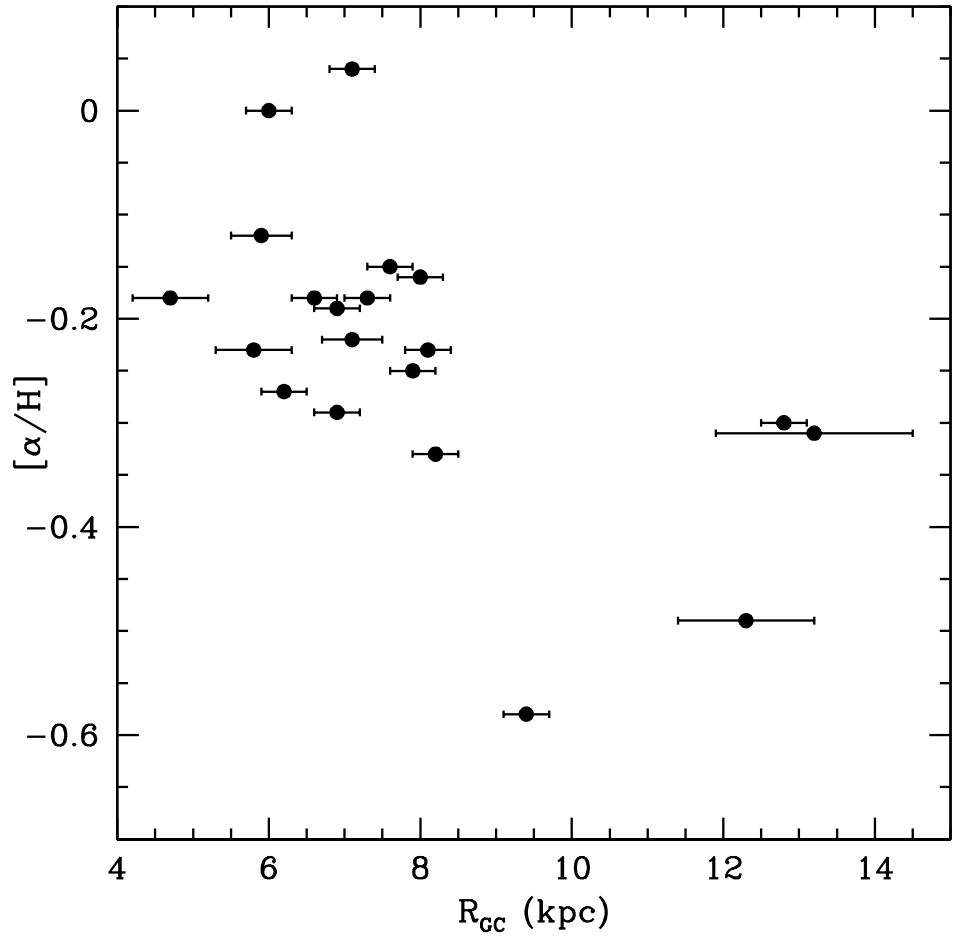}
\caption{The mean values of O, Mg, Si, and S relative
to solar abundances for B stars studied by Daflon \&
Cunha (2004). \label{fig:bstars}}
\end{figure}

\clearpage

\begin{figure}
\epsscale{0.8}
\plotone{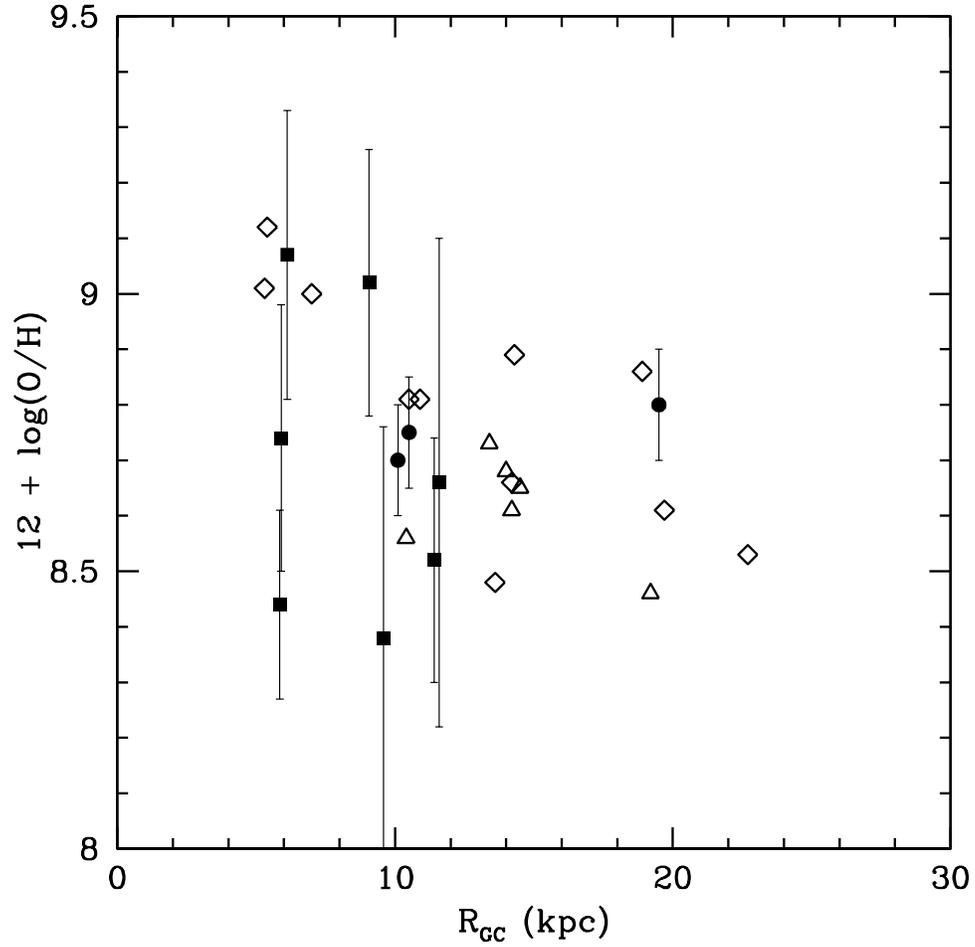}
\caption{The oxygen abundances as a function of
galactocentric distance in M31. Open triangles
are H~II regions studied by Dennefeld \& Kunth (1981), open diamonds
are H~II regions from Blair et al.\ (1982), filled circles
are stars from Venn et al.\ (2000), and filled squares are
stars from Trundle et al.\ (2002).
\label{fig:m31otoh}}
\end{figure}

\clearpage

\begin{figure}
\epsscale{0.8}
\plotone{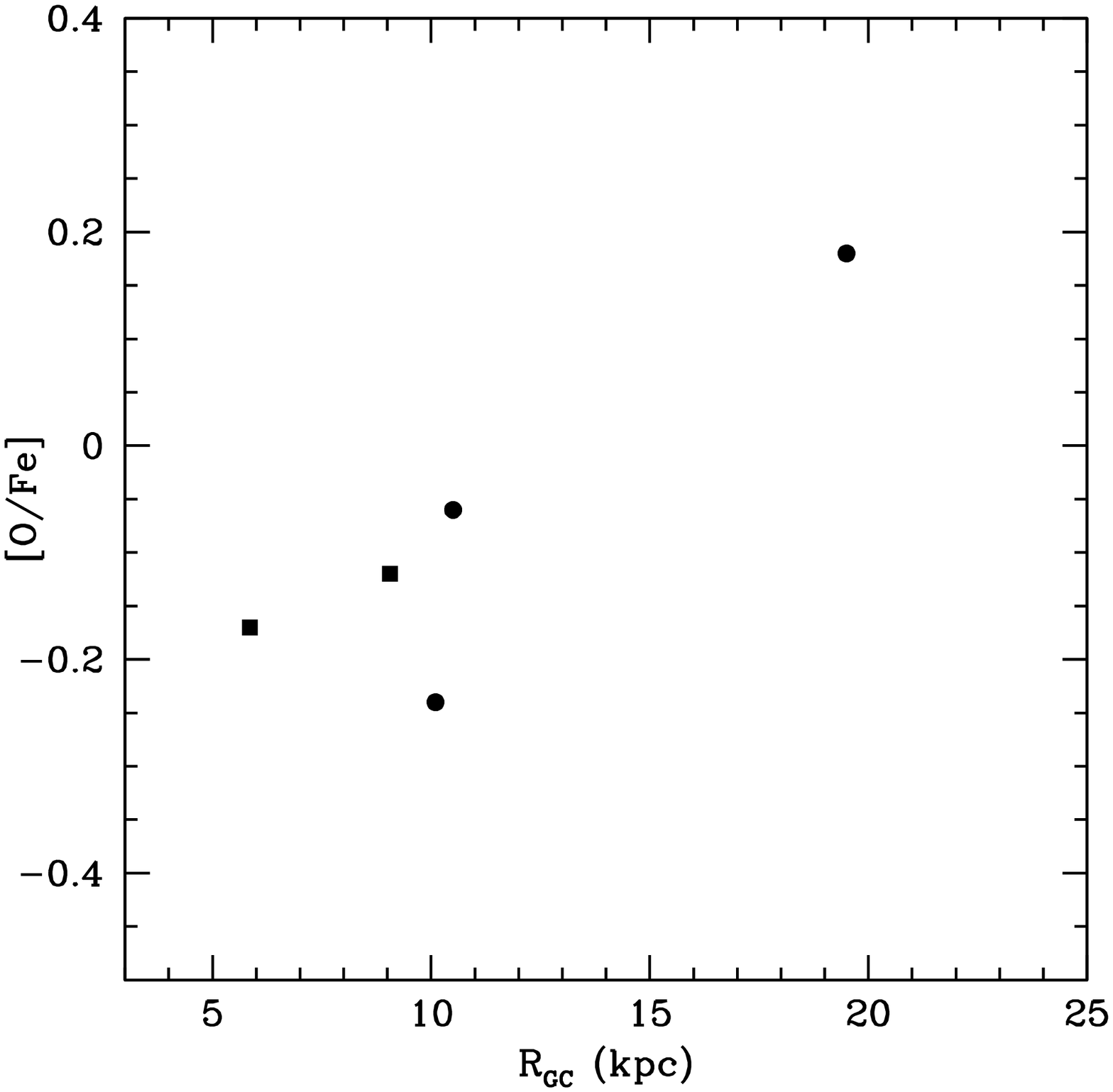}
\caption{The oxygen-to-iron ratio of A and B
supergiants in M31, with the same symbols as
Figure~\ref{fig:m31otoh}.
\label{fig:m31ofe}}
\end{figure}


\clearpage

\begin{references}

\reference{} Alonso, A., Arribas, S., \& Martinez Roger, C.\ 1994, \aaps, 107, 365

\reference{} Alonso, A., Arribas, S., \& Martinez Roger, C.\ 1999, \aaps, 140, 261

\reference{} Alves, D.\ R.\ 2000, \apj, 539, 732

\reference{} Andrievsky, S.\ M., Bersier, D., Kovtyukh, V.\ V., Luck, R.\ E.,
Maciel, W.\ J., L\'{e}pine, J.\ R.\ D., \& Beletsky, Yu.\ V.\ 2002a, \aap, 384, 140

\reference{} Andrievsky, S.\ M., Kovtyukh, V.\ V., Luck, R.\ E., L\'{e}pine, J.\ R.\ D.,
Bersier, D., Maciel, W.\ J., Barbuy, B., Klochkova, V.\ G., Panchuk, V.\ E., \&
Karpischek, R.\ U.\ 2002b, \aap, 381, 32

\reference{} Andrievsky, S.\ M., Kovtyukh, V.\ V., Luck, R.\ E., L\'{e}pine, J.\ R.\ D.,
Maciel, W.\ J., \& Beletsky, Yu.\ V.\ 2002c, \aap, 392, 491

\reference{} Andrievsky, S.\ M., Luck, R.\ E., Martin, P., \& L\'{e}pine, J.\ R.\ D.\
2003, \aap, 413, 159

\reference{} Asplund, M., Grevesse, N., Sauval, A.\ J., Allende Prieto, C., \&
Kiselman, D.\ 2004, \aap, 417, 751

\reference{} Asplund, M., Nordlund, \AA., Trampedach, R., \&
Stein, R.\ F.\ 2000, \aap, 359, 743

\reference{} Bensby, T., Feltzing, S., \& Lundstr\"{o}m, I.\ 2003, \aap, 410, 527

\reference{} Bensby, T., Feltzing, S., \& Lundstr\"{o}m, I.\ 2004, \aap, 421, 969

\reference{} Bi\'{e}mont, E., Baudoux, M., Kurucz, R.\ L., Ansbacher, W.,
\& Pinnington, E.\ H.\ 1991, \aap, 249, 539

\reference{} Blackwell, D.\ E., Booth, A.\ J., Haddock, D.\ J., Petford, A.\ D.,
\& Legget, S.\ K.\ 1986a, \mnras, 220, 549

\reference{} Blackwell, D.\ E., Booth, A.\ J., Menon, S.\ L.\ R., \& Petford, A.\ D.,
1986b, \mnras, 220, 289

\reference{} Blackwell, D.\ E., Ibbetson, P.\ A., Petford, A.\ D.,
\& Shallis, M.\ J.\ 1979a, \mnras, 186, 633

\reference{} Blackwell, D.\ E., Lynas-Gray, A.\ E., \& Smith, G.\ 1995,
\aap, 296, 217

\reference{} Blackwell, D.\ E., Menon, S.\ L.\ R., \& Petford, A.\ D.\
1983, \mnras, 204, 883

\reference{} Blackwell, D.\ E., Menon, S.\ L.\ R., Petford, A.\ D.,
\& Shallis, M.\ J.\ 1982a, \mnras, 201, 611

\reference{} Blackwell, D.\ E., Petford, A.\ D., \& Shallis, M.\ J.\ 1979b, 
\mnras, 186, 657

\reference{} Blackwell, D.\ E., Petford, A.\ D., Shallis, M.\ J.,
\& Legget, S.\ 1982b, \mnras, 199, 21

\reference{} Blackwell, D.\ E., Petford, A.\ D., Shallis, M.\ J.,
\& Simmons, G.\ J.\ 1980, \mnras, 191, 445

\reference{} Blair, W.\ P., Kirshner, R.\ P., \& Chevalier, R.\ A.\ 1982,
\apj, 254, 50

\reference{} Brewer, M.-M., \& Carney, B.\ W.\ 2004, \pasa, 21, 134

\reference{} Brewer, M.-M., \& Carney, B.\ W.\ 2005, \aj, submitted

\reference{} Burton, W.\ B., \& Te Lintel Hekkert, P.\ 1986, \aap, 65, 427

\reference{} Carney, B.\ W., \& Seitzer, P.\ 1993, \aj, 105, 2127

\reference{} Carpenter, J.\ M.\ 2001, \aj, 121, 2851

\reference{} Chen, L.\ Hou, J.\ L., \& Wang, J.\ J.\ 2003, \aj, 125, 1397

\reference{} Crane, J.\ D., Majewski, S.\ R., Rocha-Pinto, H.\ J.,
Frinchaboy, P.\ M., Skrutskie, M.\ F., \& Law, D.\ R.\ 2003, \apjl, 594, L119

\reference{} Daflon, S., \& Cunha, K.\ 2004, \apj, 617, 1115

\reference{} Dennefeld, M., \& Kunth, D.\ 1981, \aj, 86, 989

\reference{} Diplas, A., \& Savage, B.\ D.\ 1991, \apj, 377, 126

\reference{} Djorgovski, S., \& Sosin, C.\ 1989, \apjl, 341, L13

\reference{} Fich, M., \& Silkey, M.\ 1991, \apj, 366, 107

\reference{} Freudenreich, H.\ T., Beriiman, G.\ B., Dwek, I., Hauser, M.\ G.,
Kelsall, T., Moseley, S.\ H., Silverberg, R.\ F., Sodroski, T.\ J., Toller,
G.\ N., \& Weiland, J.\ L.\ 1994, \apjl, 429, L69

\reference{} Frinchaboy, P.\ M., Majewski, S.\ R., Crane, J.\ D., Reid, I.\ N.,
Rocha-Pinto, H.\ J., Phelps, R.\ L., Patterson, R.\ J., \&
Mu\~{n}oz, R.\ R.\ 2004, \apjl, 602, L21

\reference{} Fuhrmann, K.\ 1998, \aap, 338, 161

\reference{} Graham, J.\ A.\ 1982, \pasp, 94, 244

\reference{} Grevesse, N., \& Sauval, A.\ J.\ 1998, Space Sci.\ Reviews, 85, 161

\reference{} Henderson, A.\ P., Jackson, P.\ D., \& Kerr, F.\ J.\ 1982,
\apj, 263, 116

\reference{} Henry, R.\ B.\ C., \& Worthey, G.\ 1999, \pasp, 111, 919

\reference{} Ibata, R.\ A., Irwin, M.\ J., Lewis, G.\ F., Ferguson, A.\ M.\ N.,
\& Tanvir, N.\ 2003, \mnras, 340, L21

\reference{} Jacoby, G.\ H., \& Ciardullo, R.\ 1999, \apj, 515, 169

\reference{} Janes, K.\ A.\ 1979, \apjs, 39, 135

\reference{} Kurucz, R.\ L., \& Bell, B.\ 1995, Kurucz CD-ROM No.\ 23: Smithsonian
Astrophysical Observatory

\reference{} Lambert, D.\ L., \& Luck, R.\ E.\ 1976, The Observatory, 96, 100

\reference{} Lawler, J.\ E., Bonvallet, G., \& Sneden, C.\ 2001a, \apj, 556, 452

\reference{} Lawler, J.\ E., Wickliffe, M.\ E., den Hartog, E.\ A., \&
Sneden, C.\ 2001b, \apj, 563, 1075

\reference{} L\'{o}pez-Corredoira, M., Cabrera-Lavers, A., Garz\'{o}n, F.,
\& Hammersley, P.\ L.\ 2002, \aap, 394, 883

\reference{} Luck, R.\ E., Gieren, W.\ P., Andrievsky, S.\ M., Kovtyukh, V.\ V.,
Fouqu\'{e}, P., Pont, F., \& Kienzle, F.\ 2003, \aap, 401, 939

\reference{} Martin, N.\ F., Ibata, R.\ A., Bellazzini, M., Irwin, M.\ J.,
Lewis, G.\ F., \& Dehnen, W.\ 2004, \mnras, 348, 12

\reference{} Miyamoto, M., Yoshizawa, M., \& Suzuki, S.\ 1988, \aap, 194, 107

\reference{} Newberg, H.\ J., Yanny, B., Rockosi, C.,
Grebel, E.\ K., Rix, H.-H., Brinkmann, J., Csabai, I., Hennessy, G.,
Hindsley, R.\ B., Ibata, R., Ivez\`{i}\'{c}, Z., Lamb, D., Nash, E.\ T.,
Odenkirchen, M., Rave, H.\ A., Schneider, D.\ P., Smyth, J.\ A., Stolte, A.,
\& York, D.\ G.\ 2002, \apj, 569, 245

\reference{} Orsatti, A.\ M.\ 1992, \aj, 104, 590

\reference{} Pagel, B.\ E.\ J., \& Edmunds, M.\ G.\ 1981, \araa, 19, 77

Paulson, D.\ B., Sneden, C., \& Cochran, W.\ D.\ 2003, \aj, 125, 318

\reference{} Pe\~{n}arrubia, J., Mart\'{i}nez-Delgado, D., Rix, H.\ W.,
G\'{o}mez-Flechoso, M.\ A., Munn, J., Newberg, H., Bell, E.\ F.,
Yanny, B., Zucker, D., \& Grebel, E.\ K.\ 2004, astro-ph/0410448

\reference{} Prochaska, J.\ X., \& McWilliam, A.\ 2000, \apj, 537, L57

\reference{} Prochaska, J.\ X., Naumov, S.\ O., Carney, B.\ W.,
McWilliam, A., \& Wolfe, A.\ M.\ 2000, \aj, 120, 2513

\reference{} Ram\'{i}rez, S.\ V., \& Cohen, J.\ G.\ 2002, \aj, 123, 3277

\reference{} Reed, B.\ C.\ 1996, \aj, 111, 804

\reference{} Rocha-Pinto, H.\ J., Majewski, S.\ R., Skrutskie, M.\ F., \&
Crane, J.\ D.\ 2003, \apjl, 594, L115

\reference{} Rolleston, W.\ R.\ J., Smartt, S.\ J., Dufton, P.\ L., \&
Ryans, R.\ S.\ I.\ 2000, \aap, 363, 537

\reference{} Schlegel, D.\ J., Finkbeiner, D.\ P., \& Davis, M.\
1998, \apj, 500, 525

\reference{} Shaver, P., McGee, R.\ X., Newton, L.\ M., Danks, A.\ C.,
\& Pottasch, S.\ R.\ 1983, \mnras, 204, 53

\reference{} Smith, G., \& Raggett, D.\ St.\ J.\ 1981, J.\ Phys.\ B, 14, 4015

\reference{} Sneden, C.\ 1973, \apj, 184, 839

\reference{} Sodroski, T.\ J., Dwek, E., Hauser, M.\ G., \& Kerr, F.\ J.\ 1987,
\apj, 322, 101

\reference{} Thor\'{e}n, P., \& Feltzing, S.\ 2000, \aap, 393, 692

\reference{} Tomkin, J., \& Lambert, D.\ L.\ 1999, \apj, 523, 234

\reference{} Trundle, C., Dufton, P.\ L., Lennon, D.\ J.,
Smartt, S.\ J., \& Urbaneja, M.\ A.\ 2002, \aap, 395, 519

\reference{} Twarog, B.\ A., Ashman, K.\ M., \& Anthony-Twarog, B.\ J.\ 1997,
\aj, 114, 2556

\reference{} Venn, K.\ A., Irwin, M., Shetrone, M.\ D., Tout, C.\ A., Hill, V.,
\& Tolstoy, E.\ 2004, \aj, 128, 1177

\reference{} Venn, K.\ A., McCarthy, J.\ K.,. Lennon, D.\ J., Przybilla, N.,
Kudritzki, R.\ P., \& Lemke, M.\ 2000, \apj, 541, 510

\reference{} V\'{i}lchez, J.\ M., \& Esteban, C.\ 1996, \mnras, 280, 720

\reference{} Wouterloot, J.\ G.\ A., Brand, J., Burton, W.\ B., \&
Kwee, K.\ K.\ 1990, \aap, 230, 21

\reference{} Yanny, B., Newberg, H.\ J., 
Grebel, E.\ K., Kent, R., Odenkirchen, M., Rockosi, C.\ M., Schlegel, D.,
Subbarao, M., Brinkmann, J., Fukugita, M.,
Ivez\`{i}\'{c}, Z., Lamb, D., Schneider, D.\ P.,
\& York, D.\ G.\ 2003, \apj, 588, 824

\reference{} Yong, D., Carney, B.\ W., \& de~Almeida, M.\ L.\ T.\ 
2005, \aj, submitted

\reference{} Zaritsky, D., Kennicutt, Jr., R.\ C., \& Huchra, J.\ P.\ 1994,
\apj, 420, 87

\end{references}
\end{document}